\documentclass[useAMS, usegraphicx, usenatbib]{mn2e}
\usepackage{epsf}
\title[2dFGRS: correlation with REFLEX]{The 2dF Galaxy Redshift Survey: correlation with the ROSAT-ESO Flux Limited X-ray (REFLEX) galaxy cluster survey}
\author[M. Hilton et al.]{Matt Hilton,$^{1}$ Chris Collins,$^{1}$ Roberto De Propris,$^{2,3}$ Ivan K. Baldry,$^{4}$ \newauthor Carlton M. Baugh,$^{5}$ Joss Bland-Hawthorn,$^{6}$ Terry Bridges,$^{7}$ Russell Cannon,$^{6}$ \newauthor Shaun Cole,$^{5}$ Matthew Colless,$^{6}$ Warrick J. Couch,$^{8}$ Gavin B. Dalton,$^{9,10}$ \newauthor Simon P. Driver,$^{2}$ George Efstathiou,$^{11}$ Richard S. Ellis,$^{12}$ Carlos S. Frenk,$^{5}$ \newauthor Karl Glazebrook,$^{4}$ Carole A. Jackson,$^{13}$ Ofer Lahav,$^{14}$ Ian Lewis,$^{9}$ \newauthor Stuart Lumsden,$^{15}$ Steve J. Maddox,$^{16}$ Darren Madgwick,$^{17}$ Peder Norberg,$^{18}$ \newauthor John A. Peacock,$^{19}$ Bruce A. Peterson,$^{2}$ Will Sutherland$^{19}$ and Keith Taylor$^{12}$  \\
$^{1}$Astrophysics Research Institute, Liverpool John Moores University, Twelve Quays House, Egerton Wharf, Birkenhead, CH41 1LD, UK\\
$^{2}$Research School of Astronomy \& Astrophysics, The Australian National University, Weston Creek, ACT 2611, Australia\\
$^{3}$Astrophysics Group, Department of Physics, University of Bristol, Tyndall Avenue, Bristol, BS8 1TL, UK\\
$^{4}$Department of Physics \& Astronomy, Johns Hopkins University, Baltimore, MD 21118-2686, USA\\
$^{5}$Department of Physics, University of Durham, South Road, Durham, DH1 3LE, UK\\
$^{6}$Anglo-Australian Observatory, P.O. Box 296, Epping, NSW 2111, Australia\\
$^{7}$Department of Physics, Queen's University, Kingston, Ontario, K7L 3N6, Canada\\
$^{8}$Department of Astrophysics, University of New South Wales, Sydney, NSW 2052, Australia\\
$^{9}$Department of Physics, University of Oxford, Keble Road, Oxford, OX1 3RH, UK\\
$^{10}$Space Science \& Technology Division, Rutherford Appleton Laboratory, Chilton, OX11 0QX, UK\\
$^{11}$Institute of Astronomy, University of Cambridge, Madingley Road, Cambridge, CB3 0HA, UK\\
$^{12}$Department of Astronomy, California Institute of Technology, Pasadena, CA 91025, USA\\
$^{13}$CSIRO Australia Telescope National Facility, P.O. Box 76, Epping, NSW 1710, Australia\\
$^{14}$Department of Physics \& Astronomy, University College London, Gower Street, London, WC1E 6BT, UK\\
$^{15}$Department of Physics, University of Leeds, Woodhouse Lane, Leeds, LS2 9JT, UK\\
$^{16}$School of Physics \& Astronomy, University of Nottingham, Nottingham, NG7 2RD, UK\\
$^{17}$Department of Astronomy, University of California, Berkeley, CA 94720, USA\\
$^{18}$ETHZ Institut f\"{u}r Astronomie, HPF G3.1, ETH H\"{o}nggerberg, CH-8093 Z\"{u}rich, Switzerland\\
$^{19}$Institute for Astronomy, University of Edinburgh, Royal Observatory, Blackford Hill, Edinburgh, EH9 3HJ, UK}
\begin{document}

\date{Released 2054 Xxxxx XX}

\pagerange{\pageref{firstpage}--\pageref{lastpage}} \pubyear{2054}

\label{firstpage}

\maketitle

\begin{abstract}
The ROSAT-ESO Flux Limited X-ray (REFLEX) galaxy cluster survey and the 2dF Galaxy Redshift Survey (2dFGRS) respectively comprise the largest, homogeneous X-ray selected cluster catalogue and completed galaxy redshift survey. In this work we combine these two outstanding datasets in order to study the effect of the large-scale cluster environment, as traced by X-ray luminosity, on the properties of the cluster member galaxies. We measure the $L_{\rm X}-\sigma_r$ relation from the correlated dataset and find it to be consistent with recent results found in the literature. Using a sample of 19 clusters with $L_{\rm X} \geq 0.36 \times 10^{44}$ erg s$^{-1}$ in the (0.1--2.4 keV) band, and 49 clusters with lower X-ray luminosity, we find that the fraction of early spectral type ($\eta \leq-1.4$), passively-evolving galaxies is significantly higher in the high-$L_{\rm X}$ sample within $R_{200}$. We extend the investigation to include composite $b_{\rm J}$ cluster luminosity functions, and find that the characteristic magnitude of the Schechter-function fit to the early-type luminosity function is fainter for the high-$L_{\rm X}$ sample compared to the low-$L_{\rm X}$ sample ($\Delta M^*=0.58\pm0.14$). This seems to be driven by a deficit of such galaxies with $M_{b_J}\sim-21$. In contrast, we find no significant differences between the luminosity functions of star-forming, late-type galaxies. We believe these results are consistent with a scenario in which the high-$L_{\rm X}$ clusters are more dynamically evolved systems than the low-$L_{\rm X}$ clusters.

\end{abstract}


\section{Introduction}
It has been known since the earliest observations of rich clusters \citep[see e.g.,][]{Abell_1965} that galaxies which inhabit these high-density regions are quite distinct from the general field galaxy population. The cluster population is dominated by galaxies with early-type morphologies, primarily ellipticals and S0s, which have colours and spectral types consistent with their undergoing passive evolution. In contrast, the field population is dominated by actively star-forming galaxies generally of late morphological types, such as spirals and irregulars. Explaining the segregation of galaxies into these two broad classes is one of the outstanding problems of extragalactic astronomy and cosmology. 

Early work suggested that elliptical galaxies may have an intrinsically different formation process to disc galaxies \citep[e.g.,][]{Sandage_1970, GottThuan_1976}. However, since the general acceptance of the hierarchical process as the preferred model of structure formation \citep[see e.g.,][]{Kauffmann_1993, Baugh_1996, Cole_2000}, much attention has been focused on mechanisms that could transform star-forming, late-type galaxies into passively evolving, early-type systems in dense environments. Some examples of suggested evolutionary processes include: galaxy mergers -- the results of numerical simulations suggest that the collision and merging of two equal mass disc galaxies can produce a product with properties typical of elliptical galaxies \citep[e.g.,][]{Barnes_1992}; ram-pressure stripping of gas from galaxies falling into clusters \citep{GunnGott_1972}; and galaxy harassment, where tidal forces strip disc galaxies to make dwarf spheroidal systems \citep{MooreHarassment_1999}.

Modern large surveys such as the 2dF Galaxy Redshift Survey \citep[2dFGRS, ][]{Colless2dF_2001} and the Sloan Digital Sky Survey \citep[SDSS, ][]{YorkSDSS_2000}, provide excellent quality data and have allowed the dependence of galaxy properties upon environment in the local universe to be examined in some depth in recent years. \citet{Lewis2dF_2002} studied the environmental dependence of the galaxy star formation rate near clusters in the 2dFGRS and found that it converged to the field value beyond $\sim3$ times the virial radius, indicating that relatively small increments in local density at the outskirts of clusters lead to a decrease in star formation rates. They found no dependence of the average star formation rate upon cluster velocity dispersion, a tracer of mass and hence a measure of the global environment. \citet{Gomez_2003} used the Early Data Release of the SDSS to study the galaxy star formation rate as functions of local galaxy density and clustercentric radius. They found that the star formation rate is strongly correlated with the local galaxy density. Similarly to \citet{Lewis2dF_2002}, \citet{Gomez_2003} also found that the star formation rate converges to the field value at clustercentric distances of 3-4 virial radii. \citet{Balogh2dFSDSS_2004} extended these studies of `galaxy ecology' to groups, using both 2dFGRS and SDSS data. They found little evidence that the distribution of the H$\alpha$ line strength, an indicator of star formation rate, depends strongly upon environment amongst the actively star-forming galaxy population. Similarly to both \citet{Lewis2dF_2002} and \citet{Gomez_2003}, they also found that the fraction of galaxies with significant ongoing star formation decreased steadily with increasing density, and interpreted the persistence of this correlation at low densities as indicating that ram pressure stripping is not the only mechanism responsible for the truncation of star formation in galaxies. In addition, \citet{Balogh2dFSDSS_2004} found little dependence of the star formation rate upon group velocity dispersion.
 
The luminosity function (LF) of galaxies is a probability distribution over absolute magnitude, and by measuring the shape of the distribution in a number of environments, one is able to obtain clues as to which processes are important in shaping galaxy evolution. \citet[][hereafter DP03]{dePropLFs_2003}, constructed $b_{\rm J}$ composite luminosity functions (LFs) of rich 2dFGRS clusters and found that the LFs of early spectral type galaxies have brighter characteristic magnitudes $M^*$ and steeper faint end slopes $\alpha$ than the 2dFGRS field LFs of \citet{MadgwickLFs_2002}. In contrast, the LFs of star-forming galaxies were found to be essentially identical. DP03 also constructed LFs for a variety of cluster subsamples, divided by several environmental variables including richness, velocity dispersion, substructure, and B--M \citep{BaulzMorgan_1970} type, but found no evidence of differences in the derived LF parameters. However, one measure of the large scale cluster environment that DP03 were unable to test was cluster X-ray emission, which is produced from the thermal bremsstrahlung radiation of the hot intracluster gas \citep[see e.g.,][]{Sarazin_1986}, and is perhaps a better tracer of cluster mass than the velocity dispersion. There have of course been many other studies of the LF in different environments in recent years; we refer the reader to the references of DP03 for details of these.

In this paper we seek to build upon the previous work described above and investigate the dependence of cluster galaxy populations upon the X-ray luminosity of the host cluster, by utilising the ROSAT-ESO Flux Limited X-ray galaxy cluster survey (REFLEX) catalogue \citep{BohringerREFLEXCat_2004} in conjunction with galaxy data taken from the 2dFGRS. This subject is also being explored by the RASS-SDSS galaxy cluster survey \citep{PopessoI_2004, PopessoII_2005, PopessoIII_2005, PopessoIVpre_2005}, who have examined cluster scaling relations and galaxy luminosity functions for a sample of clusters selected from ROSAT X-ray observations, with galaxy photometry and spectroscopy drawn from the SDSS. 

\begin{figure*}
\includegraphics[width=84mm]{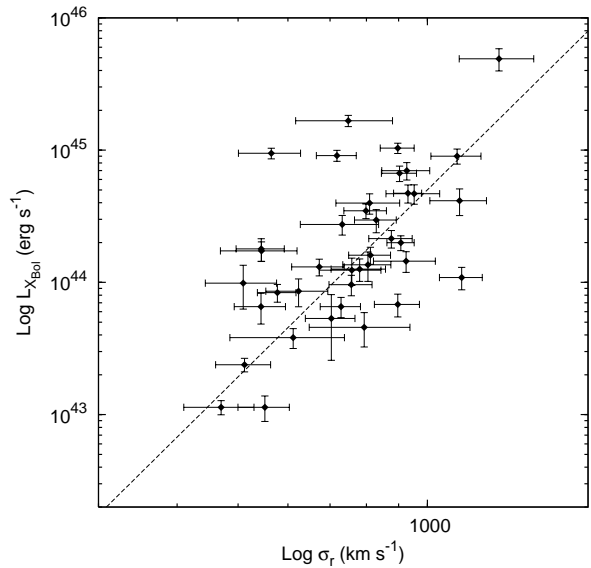}
\caption{The bolometric $L_{\rm X}-\sigma_r$ relation for the sample of 39 REFLEX clusters (dashed line). Only cluster members within $R_{\rm v}$ of the cluster centroid were used in velocity dispersion estimation.}
\label{f_LXSigma}
\end{figure*}

The outline of this paper is as follows. In the next section, we briefly describe the two datasets used in this survey, 2dFGRS and REFLEX. In Section~\ref{s_LxSBI}, we describe our initial method of correlating REFLEX and 2dFGRS data, and measure the relation between X-ray luminosity and cluster velocity dispersion. In Section~\ref{s_REFLEXnonREFLEX}, we supplement the REFLEX cluster sample with rich clusters taken from the catalogue compiled by \citet[][hereafter DP02]{dePropClusters_2002}, and divide the combined cluster sample in X-ray luminosity. We describe the construction of composite cluster luminosity functions in Section~\ref{s_LFs} for low- and high-$L_{\rm X}$ cluster subsamples, to complement the work of DP03. Finally, we end with a discussion of the results in Section~\ref{s_Discuss} and our conclusions in Section~\ref{s_Conclude}. 

Throughout we assume a concordance cosmology of $\Omega_0=0.3$, $\Omega_\Lambda=0.7$, and $H_0=70$ km s$^{-1}$ Mpc$^{-1}$.

\section{Data}
\label{s_Data}
Since the details of the 2dFGRS and REFLEX surveys have been covered in considerable depth elsewhere (see in particular \citet{Colless2dF_2001, NorbergLFs_2002} for the 2dFGRS, \citet{BohringerREFLEX_2001, Collins_2000} for REFLEX), here we only briefly review the gross characteristics of the datasets used in this paper. 

\subsection{The 2dF Galaxy Redshift Survey}
\label{s_2dFData}
The 2dFGRS is the largest completed spectroscopic survey of galaxies to date, covering approximately 1500 deg$^{2}$ of the APM galaxy survey region \citep{DaltonAPM_1997}, and contains spectroscopic information on approximately 250,000 galaxies down to a magnitude limit of $b_{\rm J}=19.45$. The survey area consists primarily of two strips: the SGP strip centred close to the South Galactic Pole, covering an area of $80^\circ \times 15^\circ$, and the NGP strip in the Northern Galactic hemisphere that covers an area of $75^\circ \times 10^\circ$. In addition there are 99 widely scattered random $2^\circ$ fields. The median redshift of the galaxies in the survey is $\bar{z}=0.11$, and up to this redshift the survey is $\sim90$ per cent complete at the stated $b_{\rm J}$ magnitude limit.

In this paper we restrict ourselves to spectra with a 2dFGRS quality flag $Q \geq 3$. These galaxies have measured redshifts with an rms uncertainty of 85 km s$^{-1}$. In our examination of galaxy populations, we make use of the 2dFGRS $\eta$ parameter defined by \citet{MadgwickEta_2003}. The value of $\eta$ is correlated with the present- to past-averaged star formation rate, and so passively-evolving galaxies have low values of $\eta$, and the opposite is true for galaxies that have formed a significant fraction of their stars more recently. In this paper we take a simple approach to classifying galaxies using this parameter, adopting $\eta \leq -1.4$ for early-type, passively evolving galaxies \citep[spectral type 1 of][]{MadgwickLFs_2002}, and $\eta >-1.4$ for late-type, actively star-forming galaxies \citep[encompassing spectral types 2-4 of][]{MadgwickLFs_2002}. Throughout this paper we adopt the following $K$-corrections for each population:

\begin{equation}
\label{e_KEarly}
K_{\rm Early}^{z}=2.6z+4.3z^2 {\quad , \rm \ and}
\end{equation}

\begin{equation}
\label{e_KLate}
K_{\rm Late}^{z}=1.3z+2.0z^2, 
\end{equation}
where $K^{z}_{\rm Early}$ is the correction found for spectral type 1 galaxies by \citet{MadgwickLFs_2002} and $K^{z}_{\rm Late}$ is that found for spectral type 3, an appropriate average to use for the late-types. The rms error on the 2dFGRS $b_{\rm J}$-band photometry is 0.15 mag.

\subsection{The ROSAT-ESO Flux Limited X-ray galaxy cluster survey}
REFLEX is the largest homogeneous catalogue of X-ray selected galaxy clusters assembled to date. REFLEX cluster candidates were selected from the region $\delta<+2.5^\circ$ of the second processing of the ROSAT All Sky Survey \citep[RASS II,][]{VogesRASSII_1999}, avoiding the crowded stellar fields of the Magellanic clouds and the region $\pm 20^\circ$ either side of the Galactic plane. REFLEX therefore covers the entire 2dFGRS region down to a nominal flux limit of $3 \times 10^{-12}$ erg s$^{-1}$ cm$^{-2}$ in the (0.1--2.4 keV) energy band. The final catalogue \citep{BohringerREFLEXCat_2004} contains 447 clusters with measured redshifts, and is $> 90$ per cent complete.

The mean fractional error on the measured REFLEX fluxes in the (0.1--2.4 keV) band is 16.7 per cent. Throughout this paper we quote REFLEX luminosities corrected for flux lost outside the detection aperture as described by \citet{BohringerNORAS_2000}. These can be found in Table 6 of \citet{BohringerREFLEXCat_2004}. 

\begin{table*}
\caption{Recent measurements of the $L_{\rm X_{Bol}}-\sigma_r$ relation, in the form $L_{\rm X_{Bol}}=10^{\alpha} \times \sigma_r{\rm (km \ s^{-1})}^{\beta} {\rm erg \ s^{-1}}$. $N_{\rm clusters}$ is the number of clusters in each sample.}
\label{t_LxSBI}
\begin{tabular}{|c|c|c|c|c|}
\hline
Reference						&	$\beta$		&	$\alpha$ 					& 	$N_{\rm clusters}$		& Comments	 \\
\hline
\citet*{WhiteJonesForman_1997}	& 5.36$\pm$0.16			& 39.3$^{+0.13}_{-0.9}$ & 14 	& Cooling flow clusters removed.\\
\citet{MahdaviGeller_2001}		& 4.4$^{+0.7}_{-0.3}$	& 31.8$^{+0.9}_{-2.0}$	& 280 	& - \\
\citet{GirardiMezzetti_2001}	& 4.4$^{+1.8}_{-1.0}$	& 29.4$^{+3.0}_{-5.4}$	& 51	& Multicomponent clusters removed.\\
\citet{XueWu_2000}				& 5.30$\pm$0.21 		& 28.32$\pm$0.61		& 197 	&	- \\
\citet{Ortiz-Gil_2004}(a)			& 4.1$\pm$0.3 			& 32.72$\pm$0.08		& 171 	& - \\
\citet{Ortiz-Gil_2004}(b)	    	& 4.2$\pm$0.4 			& 32.41$\pm$0.10		& 123  	& Multicomponent clusters removed.\\
This work						& 4.8$\pm$0.7 			& 30.6$\pm$2.1 			& 39 	& - \\
\hline
\end{tabular}
\end{table*}

\section{Correlating REFLEX and the 2\lowercase{d}FGRS}
\label{s_LxSBI}
\subsection{Membership selection within $R_{\rm v}$}
In our initial attempt at cluster membership determination, we selected 2dFGRS galaxies located within the virial radius $R_{\rm v}$ of each REFLEX cluster. We calculated $R_{\rm v}$ by adopting a self-similar model, taken from \citet{Arnaud_2002} and references therein. In the simplest self-similar assumption, $R_{\rm v}$ depends only on the cluster X-ray temperature and a fixed density contrast with respect to the critical density of the universe at redshift $z$. We converted the (0.1--2.4 keV) REFLEX X-ray luminosities to bolometric luminosities using Table 5 in the REFLEX catalogue paper \citep{BohringerREFLEXCat_2004}, the X-ray temperature of each cluster being estimated using the $L_{\rm X(0.1-2.4 \, keV)}-T$ relation of \citet{Markevitch_1998}, uncorrected for cooling flows since this was the relation used to calculate the REFLEX (0.1--2.4 keV) band X-ray luminosities. The bolometric $L_{\rm X}-T$ relation of \citet{Markevitch_1998} was then applied to estimate the X-ray temperature used in the calculation of $R_{\rm v}$ (however, in practice this extra step has a negligible effect on the estimated value of $R_v$).

Membership along the line of sight was determined in an iterative fashion, using the biweight scale estimator for velocity dispersion $\sigma_r$ recommended by \citet{BeersBiweight_1990}. Initially, $\sigma_r$ was calculated for galaxies within $\pm 2000$ km s$^{-1}$ of each REFLEX cluster redshift. On subsequent iterations, a conservative $3 \sigma_r$ clipping was applied, and galaxies with line of sight velocities outside of this range with respect to the cluster redshift were discarded. The procedure was found to converge within a few iterations, and at the end of the process clusters with less than 15 members were discarded, since the calculated velocity dispersions in these cases were unlikely to reflect the true values. The uncertainty in $\sigma_r$ was estimated via bootstrap resampling, and was found to be typically $\sim80$ km s$^{-1}$. In total, we extracted 39 REFLEX clusters with $\geq 15$ members from the 2dFGRS using this method. Typically, each REFLEX cluster contains $\sim 80$ 2dFGRS member galaxies.

\subsection{The \boldmath{$L_{\rm X}-\sigma_{\lowercase{r}}$} relation from 2\lowercase{d}F-REFLEX data}
Since the X-ray emitting gas and galaxies that make up a cluster share a common potential well, a relationship exists between $L_{\rm X}$ and $\sigma_r$. Self-similar models, where it is assumed that the evolution of clusters is solely due to their collapse under gravity, predict that $L_{\rm X_{Bol}} \propto \sigma_r^4$ \citep[see e.g.][]{QuintanaMelnick_1982}. Some authors have obtained significantly steeper values (e.g., \citealt*{WhiteJonesForman_1997}; \citealt{XueWu_2000}), and these results have been interpreted as evidence that feedback (i.e. non-gravitational heating) processes are important in clusters. 

The most relevant recent work in our case is that of \citet{Ortiz-Gil_2004}, who measured the $L_{\rm X}-\sigma_r$ relation for the REFLEX team using a combination of literature and optical follow-up data. We measured the relation using the 39 cluster 2dFGRS-REFLEX sample, to investigate whether the homogeneity of the combined dataset could overcome the small sample size and provide competitive constraints on this relation. The results for bolometric luminosities are shown in Fig.~\ref{f_LXSigma}. We used the BCES (bivariate correlated errors and intrinsic scatter) bisector method of \citet{AkritasBershady_1996} to obtain the coefficient and power-law slope estimates of the relation. This fitting technique takes into account errors in both variables and intrinsic scatter. The following results were obtained: 

\begin{equation}
\label{e_LxSBIBol}
L_{\rm X_{Bol}}=10^{30.6\pm2.1}\sigma_r{\rm (km \ s^{-1})}^{4.8\pm0.7} {\rm erg \ s^{-1}} {\rm \quad, \ and}
\end{equation}

\begin{equation}
\label{e_LxSBIROSAT}
L_{\rm X(0.1-2.4 \ keV)}=10^{32.4\pm1.7}\sigma_r{\rm (km \ s^{-1})}^{4.0\pm0.6} {\rm erg \ s^{-1}}. 
\end{equation}

The uncertainties quoted are the 1$\sigma$ errors determined by bootstrap resampling, and their large size places only weak constraints on the relation -- these results are consistent with measurements that favour feedback, but also do not rule out simple, self-similar evolution (see Table~\ref{t_LxSBI}). This is true for many of the results listed in Table~\ref{t_LxSBI} with the exception of \citet{WhiteJonesForman_1997} and \citet{XueWu_2000}, which differ significantly from the self-similar prediction for reasons that are not clear. Table~\ref{t_LxSBI} suggests that the large amount of intrinsic scatter in this relation can only be overcome by using very large sample sizes and/or cleaning the cluster sample of objects containing cooling flows. 

\section{Supplementing the REFLEX cluster sample}
\label{s_REFLEXnonREFLEX}
\subsection{Membership selection within $R_{200}$}
\label{s_MemR200}

\begin{table*}
\caption{The high-$L_{\rm X}$ cluster sample. $N_{mem}$ is the number of 2dFGRS cluster member galaxies with redshift quality flag $Q \geq 3$ used in the estimation of $\sigma_r$.}
\label{t_highLx}
\begin{tabular}{@{}|c|c|c|c|c|c|c|c|c|}
\hline
High-$L_{\rm X}$ & RA & Dec. & $z$	&  $\sigma_r$ & $R_{200}$ & $L_{\rm X(0.1-2.4}$ $_{\rm keV)}$	& $N_{mem}$	& Completeness\\
$(\geq0.36\times10^{44}$ erg s$^{-1}$)& (Deg. 2000)	& (Deg. 2000) &		&	(km s$^{-1}$)  & (Mpc)   & $(\times10^{44}$ erg s$^{-1})$	& &\\
\hline
A0954	&	153.4370	&	-0.1085	&	0.0950	&	780$\pm$80	&	1.9	&	0.764$\pm$0.114	&	48	&	0.86	\\
A0957	&	153.4180	&	-0.9144	&	0.0451	&	730$\pm$50	&	1.8	&	0.434$\pm$0.056	&	85	&	0.85	\\
A1650	&	194.6710	&	-1.7569	&	0.0841	&	700$\pm$60	&	1.7	&	3.873$\pm$0.236	&	97	&	0.84	\\
A1651	&	194.8390	&	-4.1948	&	0.0842	&	940$\pm$70	&	2.2	&	4.289$\pm$0.257	&	115	&	0.80	\\
A1663	&	195.7110	&	-2.5062	&	0.0829	&	810$\pm$70	&	1.9	&	0.815$\pm$0.178	&	88	&	0.84	\\
A1750	&	202.7080	&	-1.8728	&	0.0858	&	920$\pm$50	&	2.2	&	2.291$\pm$0.266	&	116	&	0.82	\\
A2734	&	2.8364	&	-28.8551	&	0.0612	&	810$\pm$60	&	2.0	&	1.197$\pm$0.108	&	108	&	0.82	\\
A2811	&	10.5362	&	-28.5358	&	0.1078	&	950$\pm$70	&	2.2	&	3.030$\pm$0.297	&	94	&	0.93	\\
A3027	&	37.6814	&	-33.0987	&	0.0774	&	910$\pm$70	&	2.2	&	0.449$\pm$0.070	&	90	&	0.75	\\
A3094	&	47.8540	&	-26.8998	&	0.0683	&	710$\pm$60	&	1.7	&	0.362$\pm$0.182	&	90	&	0.90	\\
A3880	&	336.9690	&	-30.5699	&	0.0577	&	820$\pm$60	&	2.0	&	0.939$\pm$0.077	&	105	&	0.96	\\
A4038	&	356.9300	&	-28.1414	&	0.0303	&	910$\pm$40	&	2.2	&	1.127$\pm$0.043	&	155	&	0.91	\\
S1136	&	354.0710	&	-31.6103	&	0.0620	&	650$\pm$70	&	1.6	&	0.549$\pm$0.113	&	43	&	0.89	\\
S0084	&	12.3502	&	-29.5244	&	0.1087	&	820$\pm$70	&	1.9	&	1.566$\pm$0.251	&	52	&	0.93	\\
S0041	&	6.3849	&	-33.0472	&	0.0494	&	580$\pm$50	&	1.4	&	0.537$\pm$0.052	&	70	&	0.81	\\
RXCJ1326.2+13	&	201.5740	&	0.2257	&	0.0827	&	560$\pm$110	&	1.3	&	1.001$\pm$0.117	&	31	&	0.81	\\
RXCJ1309.2-136	&	197.3210	&	-1.6126	&	0.0831	&	560$\pm$70	&	1.3	&	1.030$\pm$0.156	&	30	&	0.88	\\
RXCJ0229.3-3332	&	37.3430	&	-33.5378	&	0.0774	&	760$\pm$60	&	1.8	&	0.606$\pm$0.078	&	58	&	0.73	\\
RXCJ0225.1-2928	&	36.2939	&	-29.4740	&	0.0607	&	550$\pm$60	&	1.3	&	0.434$\pm$0.102	&	53	&	0.73	\\
\hline
\end{tabular}
\end{table*}

\begin{table*}
\caption{The low-$L_{\rm X}$ cluster sample. $N_{mem}$ is the number of 2dFGRS cluster member galaxies with redshift quality flag $Q \geq 3$ used in the estimation of $\sigma_r$. Clusters marked with asterisks (*) were identified by \citet{dePropClusters_2002} as being located along the same line of sight at different redshifts.}
\label{t_lowLx}
\begin{tabular}{@{}|c|c|c|c|c|c|c|c|c|}
\hline
Low-$L_{\rm X}$ & RA & Dec. & $z$	&  $\sigma_r$ & $R_{200}$ & $L_{\rm X(0.1-2.4}$ $_{\rm keV)}$	& $N_{mem}$	& Completeness\\
$(<0.36\times10^{44}$ erg s$^{-1}$)& (Deg. 2000) & (Deg. 2000) &			&	(km s$^{-1}$)  & (Mpc)   & $(\times10^{44}$ erg s$^{-1})$	& &\\
\hline
A1139	&	164.5430	&	1.5865	&	0.0396	&	550$\pm$50	&	1.3	&	0.089$\pm$0.018
	&	80	&	0.90	\\
S0301	&	42.4039	&	-31.1885	&	0.0226	&	470$\pm$60	&	1.2	&	0.089$\pm$0.008
	&	46	&	0.81	\\
MKW4	&	181.1050	&	1.9005	&	0.0200	&	510$\pm$50	&	1.3	&	0.176$\pm$0.011	&	81	&	0.96	\\
RXCJ2213.0-2753	&	333.2720	&	-27.8998	&	0.0597	&	830$\pm$170	&	2.0	&	0.316$\pm$0.085	&	24	&	0.92	\\
A1334	&	174.7647	&	-4.3176	&	0.0556	&	400$\pm$50	&	1.0	&	-	&	19	&	0.73	\\
A2660	&	356.3162	&	-25.8340	&	0.0533	&	850$\pm$60	&	2.1	&	-	&	54	&	0.73	\\
A2715*	&	0.5071	&	-34.8821	&	0.0518	&	1110$\pm$70	&	2.7	&	-	&	61	&	0.85	\\
A2716	&	0.7542	&	-27.1356	&	0.0671	&	290$\pm$50	&	0.7	&	-	&	19	&	0.85	\\
A2726	&	1.8385	&	-28.1215	&	0.0607	&	350$\pm$60	&	0.9	&	-	&	16	&	0.83	\\
A2800	&	9.5301	&	-25.0676	&	0.0637	&	400$\pm$60	&	1.0	&	-	&	25	&	0.81	\\
A2990	&	33.5477	&	-30.4606	&	0.0658	&	480$\pm$70	&	1.2	&	-	&	26	&	0.93	\\
A3095	&	48.1107	&	-27.1407	&	0.0669	&	700$\pm$60	&	1.7	&	-	&	57	&	0.77	\\
A4012	&	352.9620	&	-34.0546	&	0.0543	&	580$\pm$80	&	1.4	&	-	&	39	&	0.91	\\
A4013	&	352.5940	&	-34.9468	&	0.0535	&	490$\pm$60	&	1.2	&	-	&	52	&	0.94	\\
A4049	&	357.9031	&	-28.3647	&	0.0299	&	780$\pm$50	&	1.9	&	-	&	95	&	0.98	\\
A4049*	&	358.1844	&	-28.5704	&	0.0595	&	560$\pm$100	&	1.4	&	-	&	26	&	0.95	\\
A4053	&	358.6889	&	-27.6813	&	0.0699	&	940$\pm$60	&	2.3	&	-	&	63	&	0.88	\\
S0003	&	0.7965	&	-27.8785	&	0.0646	&	580$\pm$80	&	1.4	&	-	&	31	&	0.85	\\
S0006	&	1.1762	&	-30.4835	&	0.0285	&	550$\pm$50	&	1.3	&	-	&	29	&	0.90	\\
S0141	&	18.4462	&	-31.7481	&	0.0191	&	380$\pm$40	&	0.9	&	-	&	50	&	0.90	\\
S0160	&	22.5499	&	-32.9038	&	0.0691	&	610$\pm$100	&	1.5	&	-	&	36	&	0.89	\\
S0166	&	23.6015	&	-31.6063	&	0.0699	&	510$\pm$50	&	1.2	&	-	&	32	&	0.96	\\
S0167	&	23.5980	&	-32.8360	&	0.0658	&	760$\pm$50	&	1.8	&	-	&	53	&	0.88	\\
S0258	&	36.4352	&	-29.6160	&	0.0604	&	590$\pm$60	&	1.4	&	-	&	50	&	0.95	\\
S0333	&	48.7915	&	-29.2437	&	0.0671	&	600$\pm$60	&	1.4	&	-	&	43	&	0.88	\\
S0337	&	49.5142	&	-29.6335	&	0.0594	&	420$\pm$60	&	1.0	&	-	&	25	&	0.97	\\
S0340	&	50.0138	&	-27.0161	&	0.0671	&	980$\pm$60	&	2.4	&	-	&	52	&	1.00	\\
S1043	&	339.1166	&	-24.3418	&	0.0338	&	430$\pm$50	&	1.1	&	-	&	46	&	0.92	\\
S1155	&	357.5572	&	-29.0090	&	0.0498	&	360$\pm$70	&	0.9	&	-	&	17	&	0.95	\\
S1165	&	359.4970	&	-29.8663	&	0.0299	&	380$\pm$60	&	0.9	&	-	&	26	&	0.85	\\
S1171	&	0.3287	&	-27.4200	&	0.0277	&	330$\pm$40	&	0.8	&	-	&	22	&	0.85	\\
APMCC0809	&	340.6107	&	-24.9260	&	0.0474	&	880$\pm$210	&	2.1	&	-	&	45	&	0.85	\\
APMCC0917	&	355.3982	&	-29.2364	&	0.0510	&	540$\pm$50	&	1.3	&	-	&	61	&	0.93	\\
APMCC0945	&	359.7584	&	-31.7975	&	0.0602	&	550$\pm$60	&	1.3	&	-	&	31	&	0.96	\\
APMCC0954	&	0.2279	&	-28.4633	&	0.0615	&	490$\pm$90	&	1.2	&	-	&	31	&	0.96	\\
EDCC0069	&	329.6791	&	-28.4649	&	0.0216	&	340$\pm$50	&	0.8	&	-	&	32	&	0.82	\\
EDCC0129	&	334.7760	&	-24.1825	&	0.0355	&	1010$\pm$120	&	2.5	&	-	&	38	&	0.83	\\
EDCC0142	&	336.4215	&	-31.2005	&	0.0281	&	310$\pm$70	&	0.8	&	-	&	21	&	0.85	\\
EDCC0142*	&	336.3992	&	-31.0596	&	0.0579	&	340$\pm$50	&	0.8	&	-	&	16	&	0.85	\\
EDCC0153	&	338.0611	&	-31.2293	&	0.0576	&	520$\pm$60	&	1.3	&	-	&	30	&	0.97	\\
EDCC0155	&	338.0332	&	-25.3981	&	0.0335	&	510$\pm$60	&	1.3	&	-	&	26	&	0.79	\\
EDCC0222	&	343.8282	&	-33.9204	&	0.0287	&	290$\pm$70	&	0.7	&	-	&	28	&	0.82	\\
EDCC0321	&	354.0293	&	-32.5088	&	0.0526	&	400$\pm$100	&	1.0	&	-	&	17	&	0.89	\\
EDCC0365	&	358.7855	&	-32.7404	&	0.0592	&	510$\pm$50	&	1.2	&	-	&	44	&	0.98	\\
EDCC0442	&	6.3807	&	-33.0466	&	0.0494	&	590$\pm$50	&	1.4	&	-	&	69	&	0.81	\\
EDCC0445	&	7.1375	&	-27.5066	&	0.0618	&	590$\pm$160	&	1.4	&	-	&	16	&	0.82	\\
EDCC0457	&	9.0159	&	-26.0915	&	0.0628	&	390$\pm$50	&	0.9	&	-	&	26	&	0.95	\\
EDCC0671	&	39.2681	&	-25.3929	&	0.0570	&	390$\pm$70	&	0.9	&	-	&	17	&	0.85	\\
EDCC0697	&	42.6247	&	-34.8655	&	0.0363	&	390$\pm$40	&	0.9	&	-	&	20	&	0.80	\\
\hline
\end{tabular}
\end{table*}

Of the 39 REFLEX clusters extracted from the 2dFGRS using the method described in Section~\ref{s_LxSBI}, only 23 of these clusters are located within the redshift range $z<0.11$, for which the 2dFGRS is nearly complete, and have an overall 2dFGRS completeness (i.e. the fraction of galaxies from the parent APM galaxy catalogue with measured redshifts) at the cluster centroid of $>70$ per cent. We found that this sample was too small to practically subdivide in X-ray luminosity in order to study the effect of $L_{\rm X}$ on the cluster galaxy luminosity function. \citet[][hereafter DP02]{dePropClusters_2002} have previously studied Abell, EDCC and APMCC clusters within the 2dFGRS region, and within $z<0.11$ their catalogue should represent an essentially complete sample of rich clusters. REFLEX nominally contains all the clusters in the surveyed region above a flux limit of $3\times 10^{-12}$ erg s$^{-1}$ cm$^{-2}$ in the (0.1--2.4 keV) energy band. We therefore decided to combine the non-REFLEX clusters catalogued in DP02 with REFLEX, and use the REFLEX survey flux limit to divide the combined cluster sample in terms of X-ray luminosity.

To determine the cluster membership of the whole sample in a consistent way, we chose to replace the virial radius as used in Section~\ref{s_LxSBI} with $R_{200}$, defined as the radius within which the mean interior density is 200 times the critical density. This is straightforward to calculate by the equation \citep[from][]{CarlbergR200_1997}

\begin{equation}
R_{200}=\frac{\sqrt{3} \sigma_r}{10H(z)},
\label{e_R200}
\end{equation} 
where $H(z)$ is the Hubble constant at redshift $z$. 

Since $R_{200}$ is defined by the measured cluster velocity dispersion, it was necessary to calculate both these quantities iteratively, in step with one another. Therefore, on the first iteration the velocity dispersion was calculated using all galaxies within a 1-Mpc radius of the catalogued cluster centre. The initial velocity cut applied around the cluster redshift was $\pm 2000$ km s$^{-1}$ as before. After each iteration, $R_{200}$ was calculated using equation~(\ref{e_R200}) and the $3\sigma_r$ velocity clipping was applied. This process again converged rapidly, within 5 iterations. 

As before, clusters with less than 15 members at the end of the process were discarded. This left 36 REFLEX clusters, with measured velocity dispersions in agreement within the uncertainties of the values obtained using the selection method described in Section~\ref{s_LxSBI} (i.e. 2 clusters with few members were lost compared to using $R_{\rm v}$). $R_{200}$ is often taken to be equivalent to the virial radius, and it was found from the REFLEX sample that typically $R_{200}\sim(0.9\pm0.2)R_{\rm v}$, where the quoted uncertainty is the standard deviation. Galaxies within $R_{200}$ are therefore likely to be gravitationally bound to the cluster. 146 out of the 284 clusters catalogued by DP02 were recovered by the method. A comparison of the member numbers of the Abell clusters obtained by our method with Table 1 of DP02, who used a different method for determining cluster membership based on a `gapper' algorithm \citep[described in][]{BeersBiweight_1990}, revealed that the clusters lost in the process were located typically at $z>0.11$, and had less than the specified minimum 15 cluster members. We found generally good agreement with the velocity dispersions calculated using the biweight scale estimator in comparison to the DP02 values -- 70 per cent of the 146 clusters recovered had velocity dispersions within $<1\sigma$ of agreement. 

\subsection{Defining the cluster samples}
\label{s_sampleDefinitions}
In the following, we consider only clusters with an overall 2dFGRS redshift completeness $>70$ per cent and $z<0.11$ -- since within the survey median redshift the 2dFGRS has a very high redshift completeness up to the survey magnitude limit. These criteria result in a sample of 23 REFLEX clusters and 94 DP02 non-REFLEX clusters. 

To divide the whole cluster sample in terms of X-ray luminosity, we make use of the fact that REFLEX is a flux limited survey, assuming that the REFLEX catalogue contains all clusters above the survey flux-limit. We estimated a maximum X-ray luminosity for each non-REFLEX cluster using the nominal flux-limit of the REFLEX cluster catalogue ($3\times10^{-12}$ erg s$^{-1}$ cm$^{-2}$ in the (0.1--2.4 keV) energy band). We chose to divide the combined cluster sample into two subsamples, defined by

\begin{equation}
\label{e_lowLxSample}
L_{\rm X(0.1-2.4 \ keV)}<0.36\times10^{44} {\rm erg \ s^{-1}} {\rm \quad and}
\end{equation}

\begin{equation}
\label{e_highLxSample}
L_{\rm X(0.1-2.4 \ keV)}\geq0.36\times10^{44} {\rm erg \ s^{-1}} {\rm \ ,}
\end{equation}
for the low- and high-$L_{\rm X}$ samples respectively. The $L_{\rm X}$ value around which to divide the sample was chosen to maximise the size of the low-$L_{\rm X}$ cluster sample, while ensuring adequate coverage of the galaxy magnitude range for the construction of composite LFs for the high-$L_{\rm X}$ sample. 

The high-$L_{\rm X}$ sample consists of 19 REFLEX clusters, while the low-$L_{\rm X}$ sample is composed of 49 clusters, 4 from REFLEX with the remainder coming from the catalogue of DP02. Because of the way in which we have defined these samples, the low-$L_{\rm X}$ clusters can be thought of as essentially an optically-selected sample that has been cleaned of clusters that could potentially have X-ray luminosities above the division we have made. In the following analysis, we are therefore investigating the effect that selection of clusters based upon the observation of hot, X-ray emitting gas has on the member galaxy populations. We assume that the DP02 clusters are bound systems, since they were selected from a redshift survey and should not be artifacts of projection effects. The properties of the high- and low-$L_{\rm X}$ clusters are summarised in Tables~\ref{t_highLx} and~\ref{t_lowLx} respectively.

\begin{figure*}
\includegraphics[width=60mm]{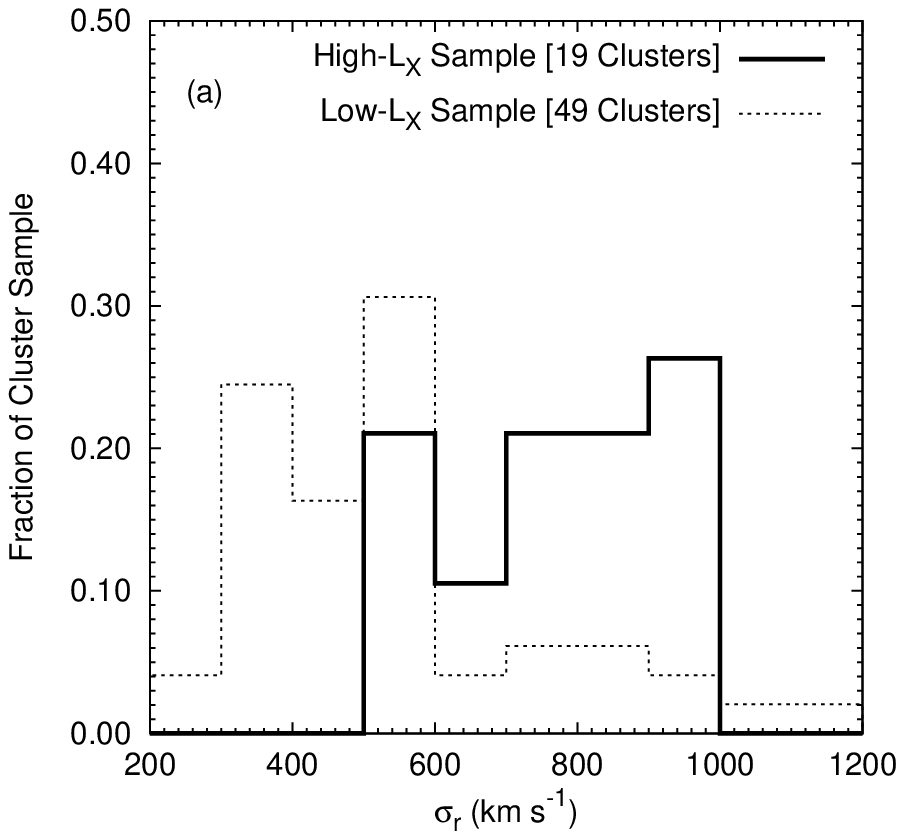}\includegraphics[width=60mm]{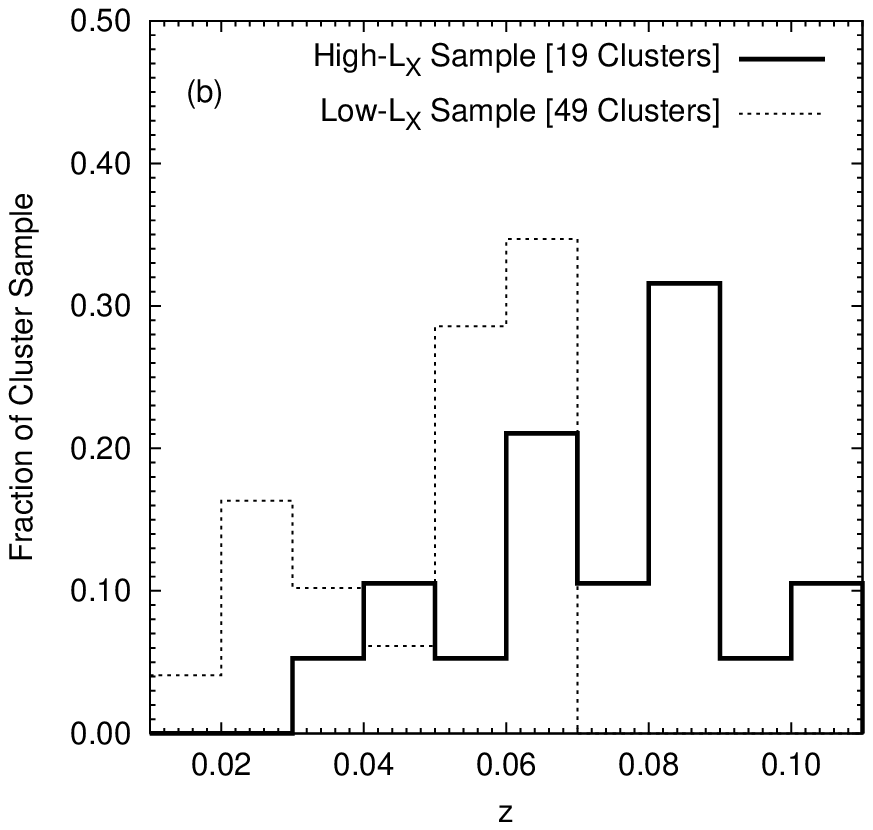}
\caption{(a) Line of sight velocity dispersion distributions for the low- and high-$L_{\rm X}$ cluster samples. (b) Redshift distributions.}
\label{f_SBIDist}
\end{figure*}

\begin{figure*}
\includegraphics[width=75mm]{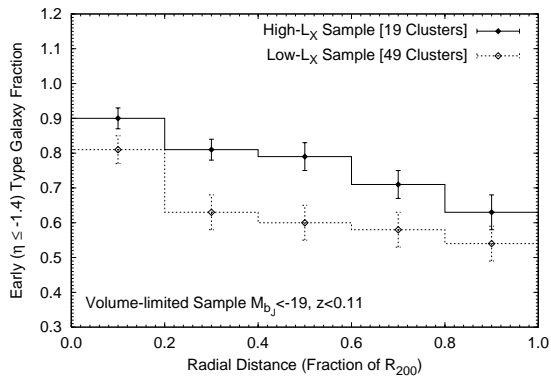}
\caption{Mean fraction of early-type ($\eta \leq -1.4$) galaxies versus fraction of $R_{200}$ for the low- and high-$L_{\rm X}$ cluster samples. A volume-limited galaxy sample, defined by $M_{b_J}<-19$, $z<0.11$, was used. The error bars are the standard error on the mean in each bin.}
\label{f_Radial}
\end{figure*}

\subsection{Properties of the cluster samples}
\label{s_properties}
Having divided the sample in terms of X-ray luminosity on the basis of several assumptions, it is useful to check the velocity dispersion distributions of the resulting low- and high-$L_{\rm X}$ samples, because $L_{\rm X}$ and $\sigma_r$ are correlated. Panel (a) of Fig.~\ref{f_SBIDist} shows that the majority of the low-$L_{\rm X}$ sample have $\sigma_r <$ 600 km s$^{-1}$, whereas in contrast most of the clusters in the high-$L_{\rm X}$ sample have $\sigma_r>700$ km s$^{-1}$. The median values of the low- and high-$L_{\rm X}$ samples were 510 km s$^{-1}$ and 780 km s$^{-1}$ respectively. This is consistent with the high-$L_{\rm X}$ sample on average containing the higher mass systems, as expected. Panel (b) of Fig.~\ref{f_SBIDist} shows the redshift distributions of both samples, and indicates that the average redshift of the high-$L_{\rm X}$ clusters is slightly higher than that for the low-$L_{\rm X}$ clusters. This is primarily due to the way in which we have defined our samples, but it is also in part due to the fact that the high-$L_{\rm X}$ clusters are intrinsically rarer systems. The median redshifts were 0.056 and 0.077 for the low- and high-$L_{\rm X}$ samples respectively. 

Under the assumption that the redshift completeness of the 2dFGRS is not a function of galaxy spectral type within $z<0.11$, we are immediately able to examine the mix of spectral types in each cluster sample. In Fig.~\ref{f_Radial}, we plot the mean fraction of passively evolving, early-type ($\eta \leq -1.4$) galaxies in composite low- and high-$L_{\rm X}$ clusters in 20 per cent radial bins of $R_{200}$. Since our samples have different median redshifts, we remove the dependence upon magnitude by using a volume-limited galaxy sample defined by $M_{b_J}<-19$, $z<0.11$. The error bars in Fig.~\ref{f_Radial} are the standard errors on the mean galaxy fraction in each bin. Clearly, the fraction of passively evolving, early-type galaxies is higher at all radii out to $R_{200}$ in the high-$L_{\rm X}$ sample compared to the low-$L_{\rm X}$ sample. In both cases, the fraction of early-type galaxies falls off smoothly with increasing radial distance from the cluster centre. The overall mean early-type galaxy fraction within $R_{200}$ is 0.76$\pm$0.02 for the high-$L_{\rm X}$ sample compared to 0.64$\pm$0.02 for the low-$L_{\rm X}$ sample.

\section{Composite luminosity functions} 
\label{s_LFs}
\subsection{Construction}

\begin{figure*}
\includegraphics[width=84mm]{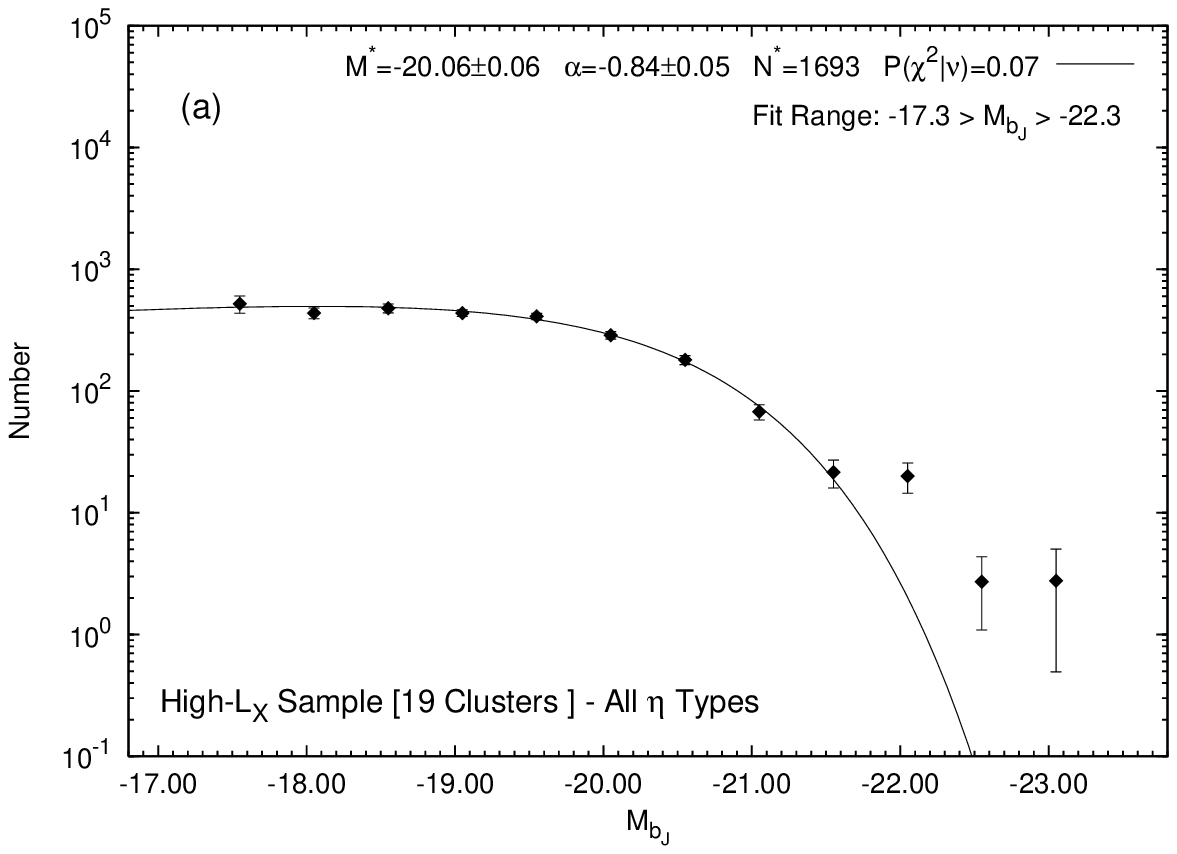}
\includegraphics[width=84mm]{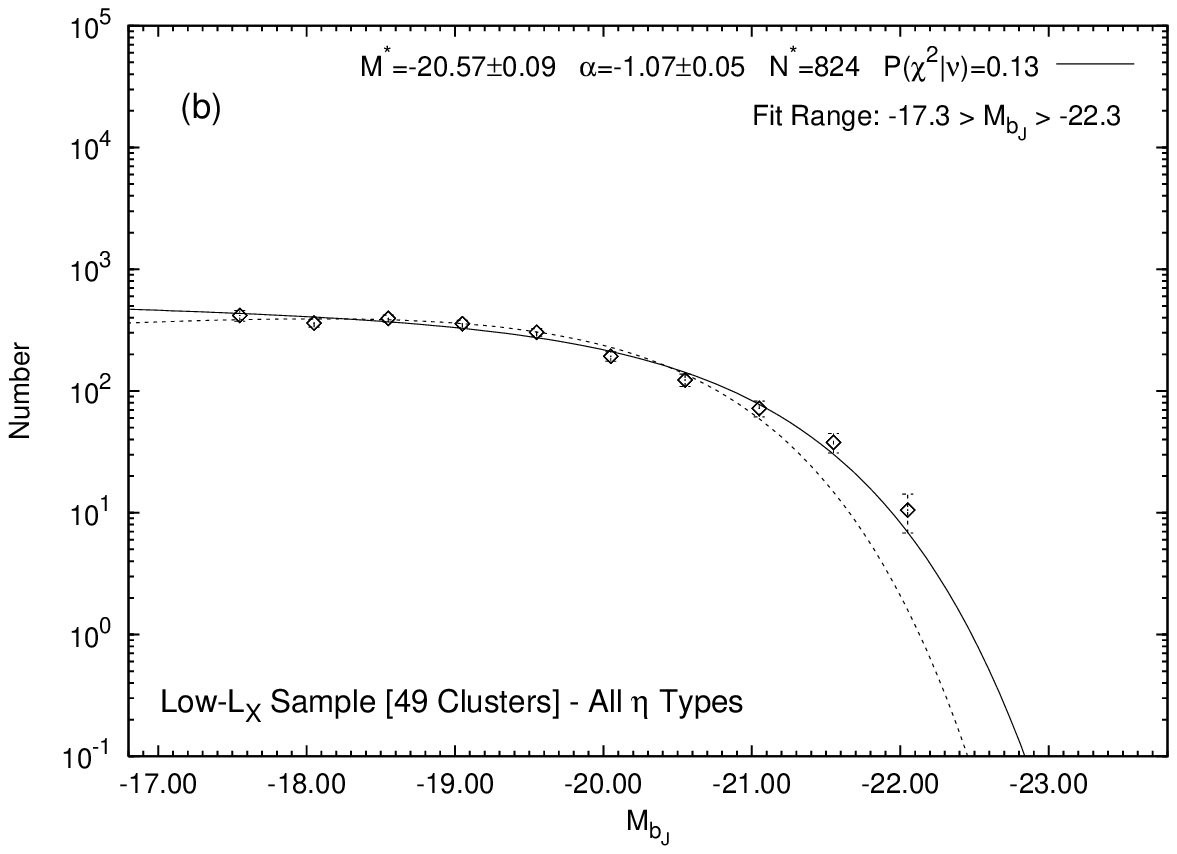}
\includegraphics[width=84mm]{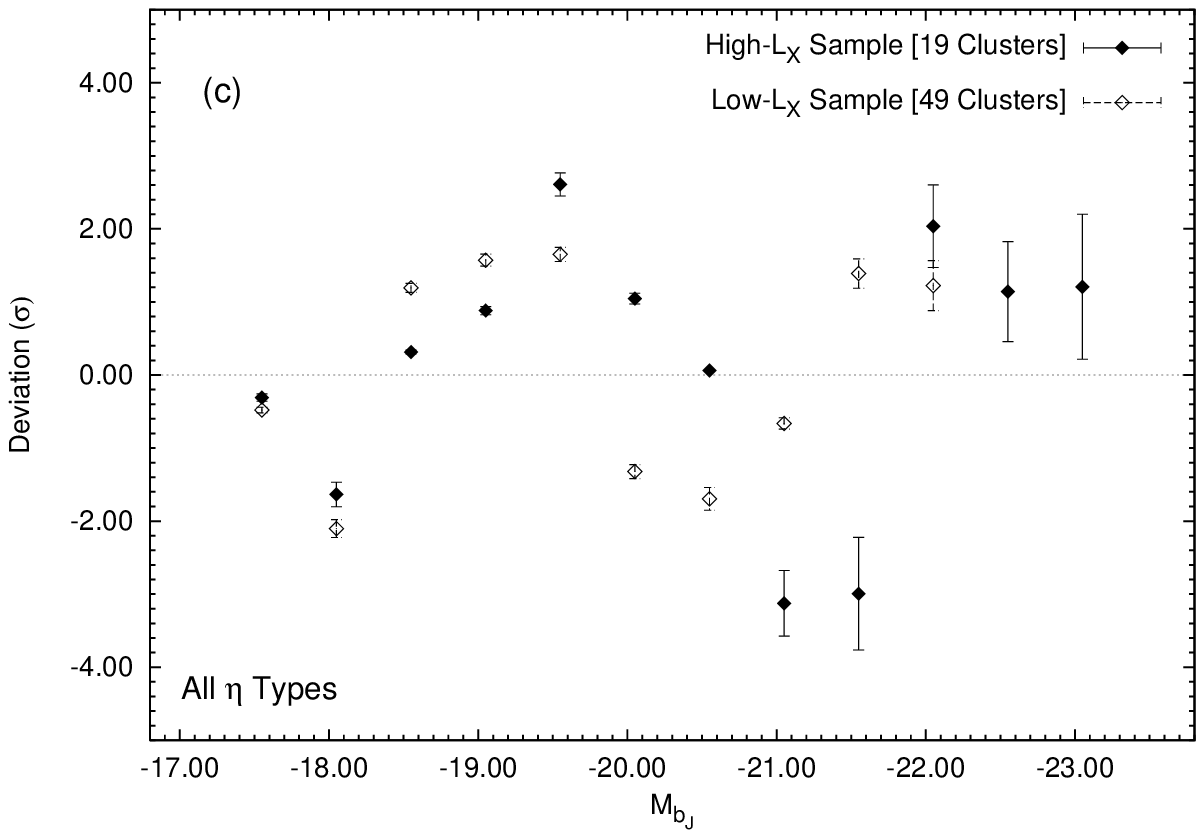}
\includegraphics[width=84mm]{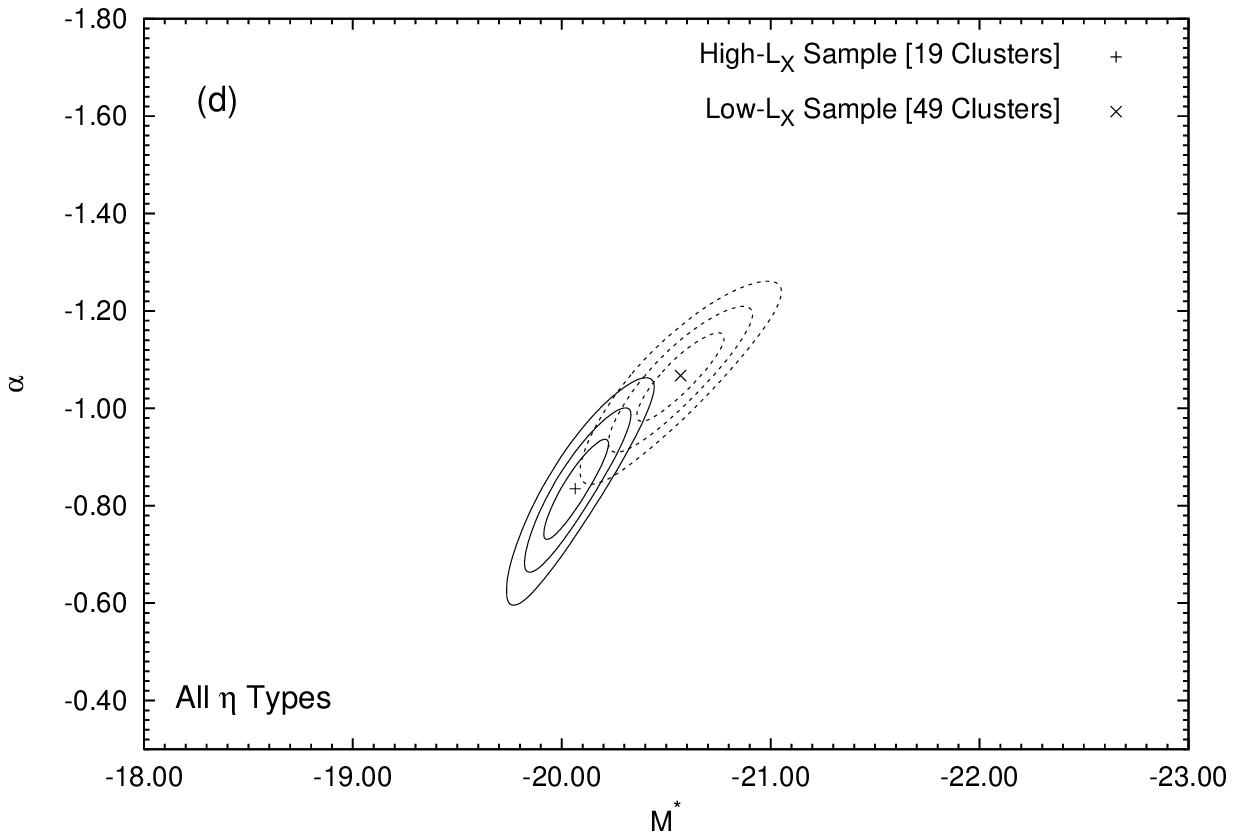}
\caption{Composite luminosity functions for all spectral types for: (a) the high-$L_{\rm X}$ sample and (b) the low-$L_{\rm X}$ sample. The solid line in each plot is the best fitting Schechter function. For comparison, we plot the renormalised fit to the high-$L_{\rm X}$ sample LF as the dashed line in panel (b). Quoted independent errors on each fitted parameter are given at the $1\sigma$ level, and were determined using a bootstrap resampling technique. (c) The deviation of each luminosity function from the best-fitting Schechter function fit to the low-$L_{\rm X}$ sample data, in units of the error on each point. (d) 1, 2, 3$\sigma$ $\chi^2$ error contours for joint luminosity function parameter estimates of the low- and high-$L_{\rm X}$ samples.}
\label{f_AllLFs}
\end{figure*}

We decided to examine the implications of the different mix of spectral types found in the low- and high-$L_{\rm X}$ clusters on the composite cluster luminosity functions by following the method of \citet{CollessLFs_1989}. The summation carried out for each absolute magnitude bin of the composite LF was

\begin{equation}
\label{e_Colless1}
N_{\rm c \it j}=\frac{N_{\rm c0}}{m_{j}}\sum_{i}\frac{N_{ij}}{N_{i0}}.
\end{equation}
Here $N_{\rm c \it j}$ is the number of galaxies in the $j$th bin of the composite LF, $N_{ij}$ is the number of galaxies contributing to the $j$th bin of the $i$th cluster LF, and $m_j$ is the number of clusters contributing to the $j$th bin of the composite LF. $N_{i0}$ is a normalisation parameter (effectively a richness) for the $i$th cluster LF, in our case calculated from the completeness corrected number of galaxies with $M_{b_{\rm J}}<-19$, which defines a volume-limited sample of 2dFGRS galaxies within $z<0.11$. $N_{\rm c0}$ is simply the sum of all the individual cluster normalisations:

\begin{equation}
\label{e_Colless2}
N_{\rm c0}=\sum_{i}N_{i0}.
\end{equation} 

In contrast to DP03, who used the parent 2dFGRS photometric catalogue (i.e. the APM galaxy survey) directly to correct for incompleteness, we corrected $N_{ij}$ for overall and magnitude-dependent incompleteness using the 2dF survey masks and software as described in Appendix A of \citet{NorbergLFs_2002}. We found that the average completeness corrections across all bins were 15 per cent for the high-$L_{\rm X}$ sample and 11 per cent for the low $L_{\rm X}$ sample. In constructing all of our LFs, we used the extinction corrected magnitudes measured by the 2dFGRS and $K$-corrected them accordingly using equations (\ref{e_KEarly}) and (\ref{e_KLate}). 

The errors in $N_{\rm c \it j}$ were calculated according to:
\begin{equation}
\label{e_Colless3}
\delta N_{\rm c \it j}=\frac{N_{\rm c0}}{m_j} \left[  \sum_i \left( \frac{\delta N_{ij}}{N_{i0}} \right) ^{2} \right] ^{1/2},
\end{equation}
where the errors in $N_{ij}$ were assumed to be Poissonian, i.e. $\sqrt{N_{ij}}$.

\subsection{Results}
\label{s_results}

\begin{figure*}
\includegraphics[width=84mm]{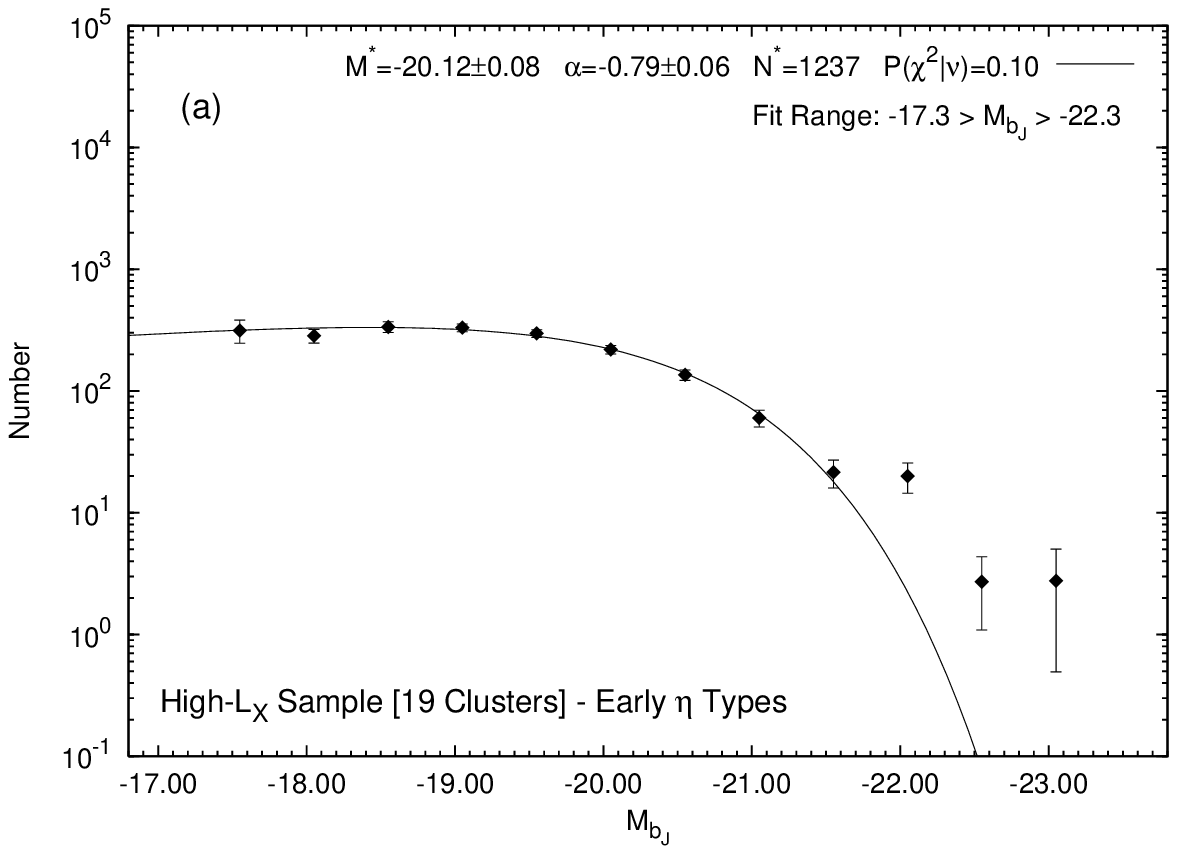}
\includegraphics[width=84mm]{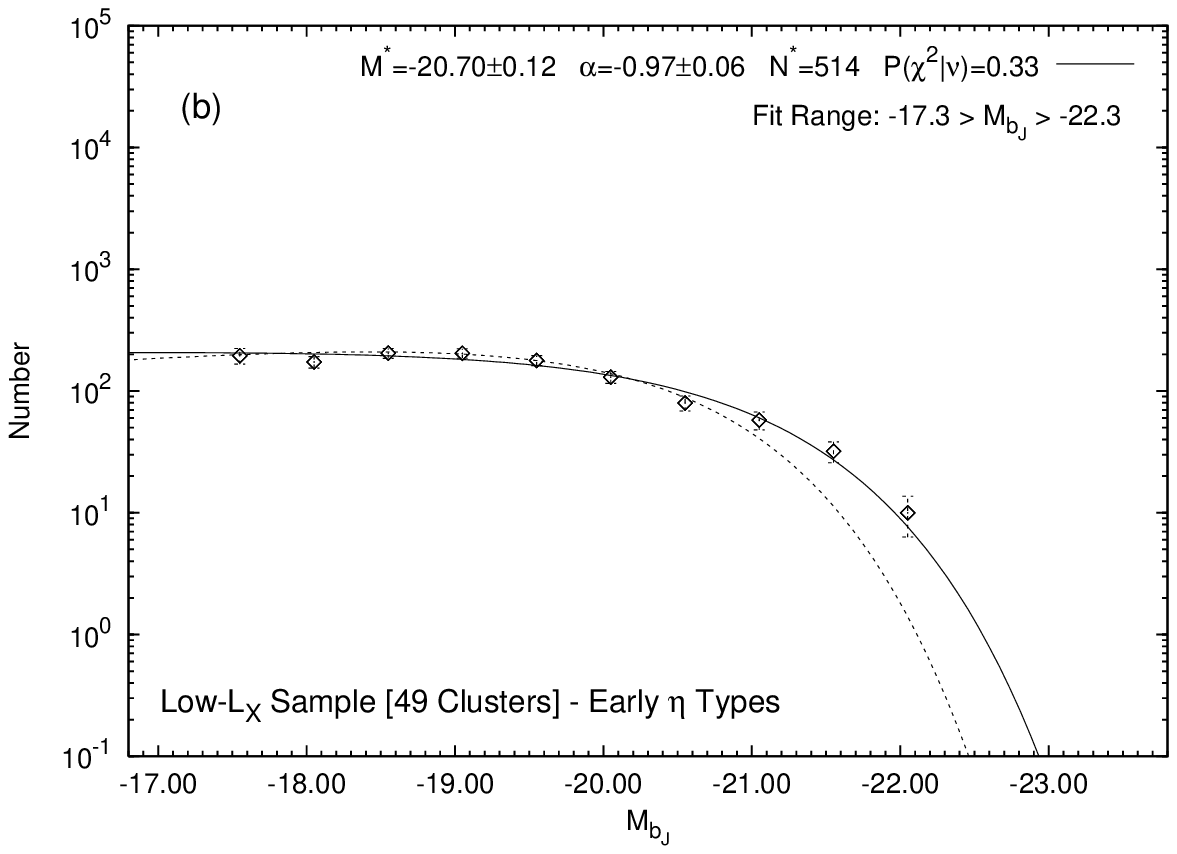}
\includegraphics[width=84mm]{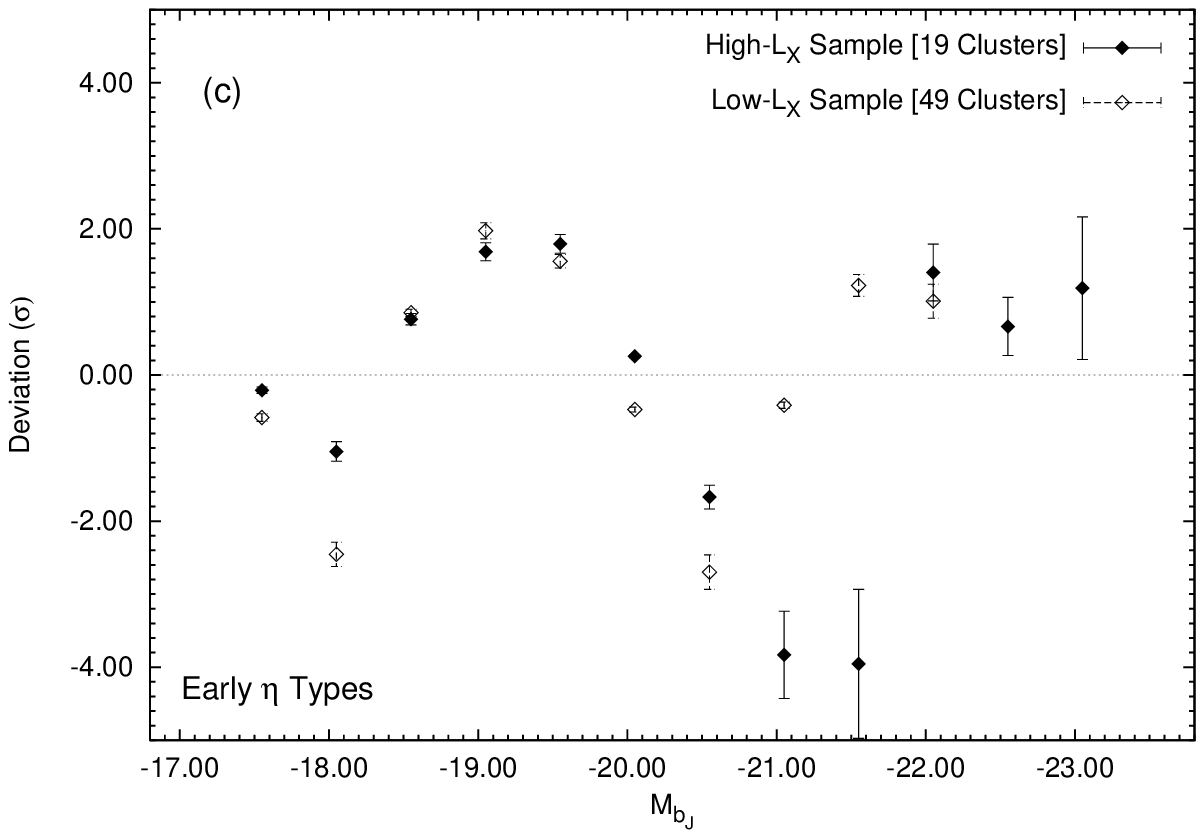}
\includegraphics[width=84mm]{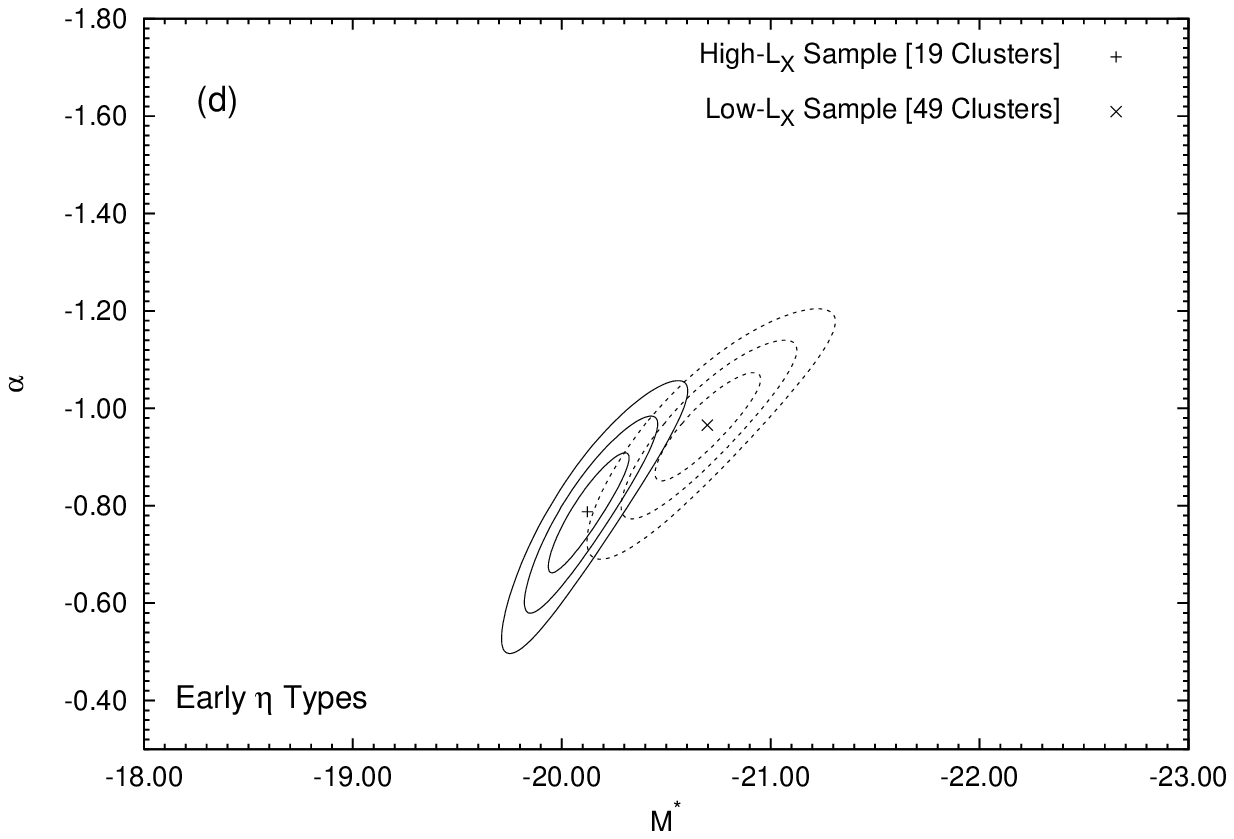}
\caption{Composite luminosity functions for early ($\eta \leq -1.4$) spectral types for: (a) the high-$L_{\rm X}$ sample and (b) the low-$L_{\rm X}$ sample. The solid line in each plot is the best fitting Schechter function. For comparison, we plot the renormalised fit to the high-$L_{\rm X}$ sample LF as the dashed line in panel (b). Quoted independent errors on each fitted parameter are given at the $1\sigma$ level, and were determined using a bootstrap resampling technique. (c) The deviation of each luminosity function from the best-fitting Schechter function fit to the low-$L_{\rm X}$ sample data, in units of the error on each point. (d) 1, 2, 3$\sigma$ $\chi^2$ error contours for joint luminosity function parameter estimates of the low- and high-$L_{\rm X}$ samples.}
\label{f_EarlyLFs}
\end{figure*}

We constructed LFs for galaxies of all, early, and late spectral types separately, for both the low- and high-$L_{\rm X}$ cluster samples. Each LF was divided into 10 bins over the magnitude range $-17.3>M_{b_{\rm J}}>-22.3$. This ensured that at least three clusters contributed to each LF bin. The faint magnitude limit that we adopt for our LFs is 0.5-mag brighter than that adopted by DP03. 

We fitted \citet{Schechter_1976} functions,

\begin{equation}
\label{e_Schechter}
n_{\rm c} \left( M \right) dM=kN^{*} e^{k \left( \alpha+1 \right) \left( M^{*}-M \right) - e^{k \left( M^{*}-M \right)}} dM,
\end{equation}
to each LF, where $M$ is the absolute $b_{\rm J}$ magnitude, $k\equiv\rm ln(10)/2.5$, $M^{*}$ is the characteristic magnitude of the turn-off at the bright-end and $\alpha$ is the faint-end slope. We performed two-parameter fits to each LF for $M^*$ and $\alpha$ using a $\chi^2$ minimisation technique. $N^*$ was fixed such that the value of the integral of the fitted function was equal to the total number of galaxies in the composite LF within the same magnitude range. We also tried fitting for the normalisation with $N^*$ held as a free parameter, but found that this made no significant difference to our results. 

\begin{figure*}
\includegraphics[width=84mm]{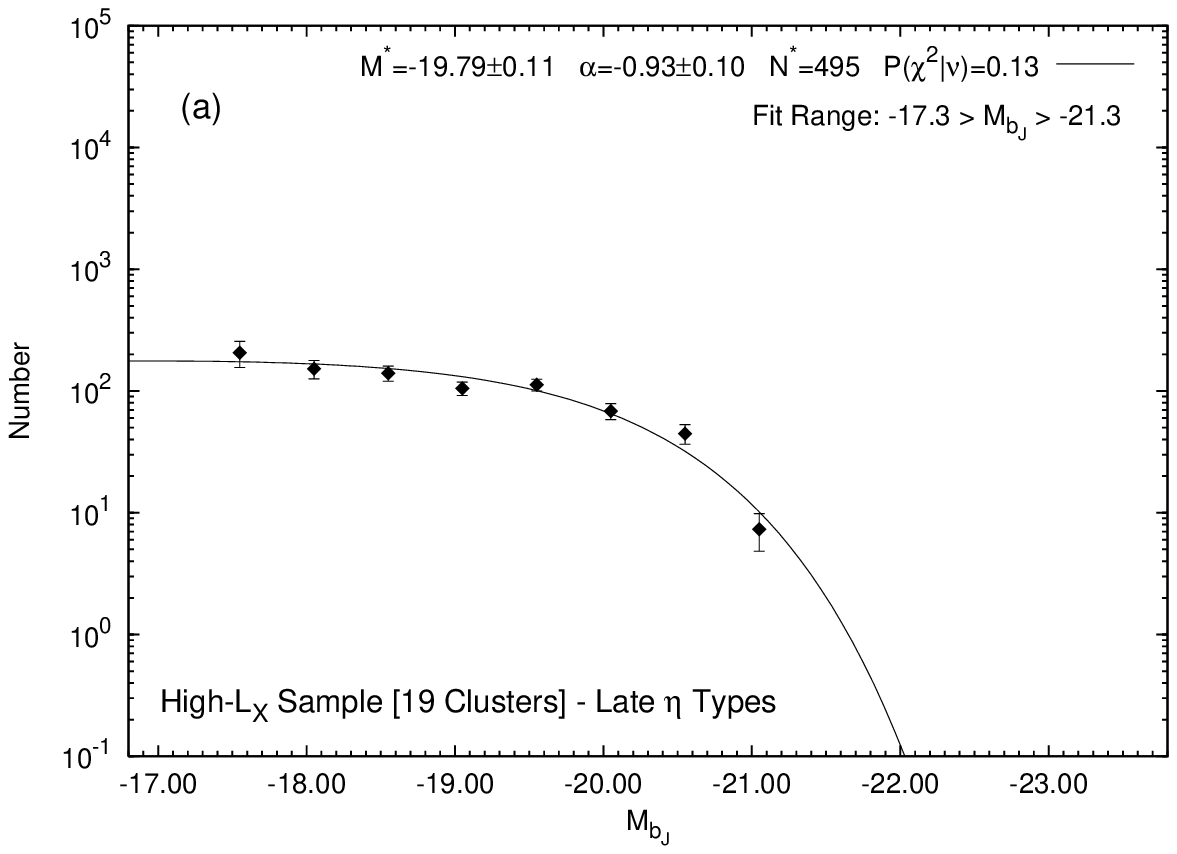}
\includegraphics[width=84mm]{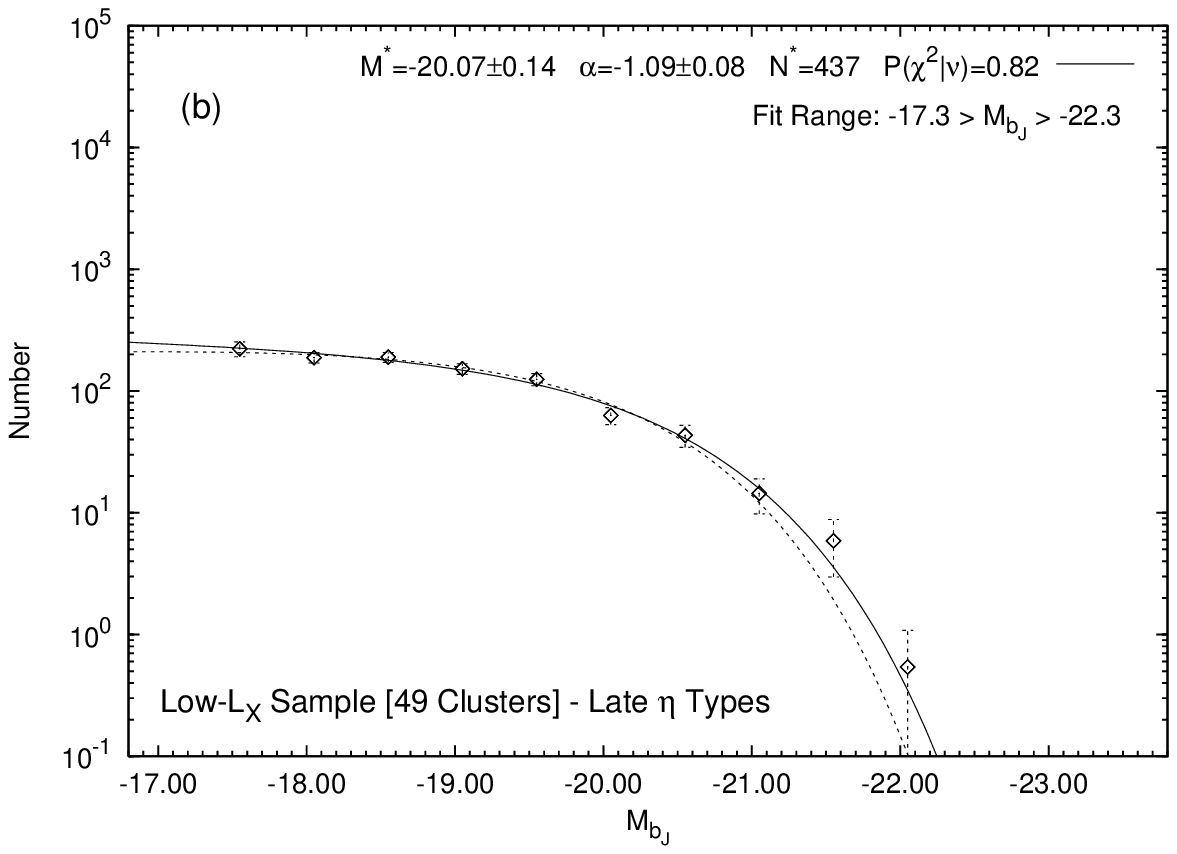}
\includegraphics[width=84mm]{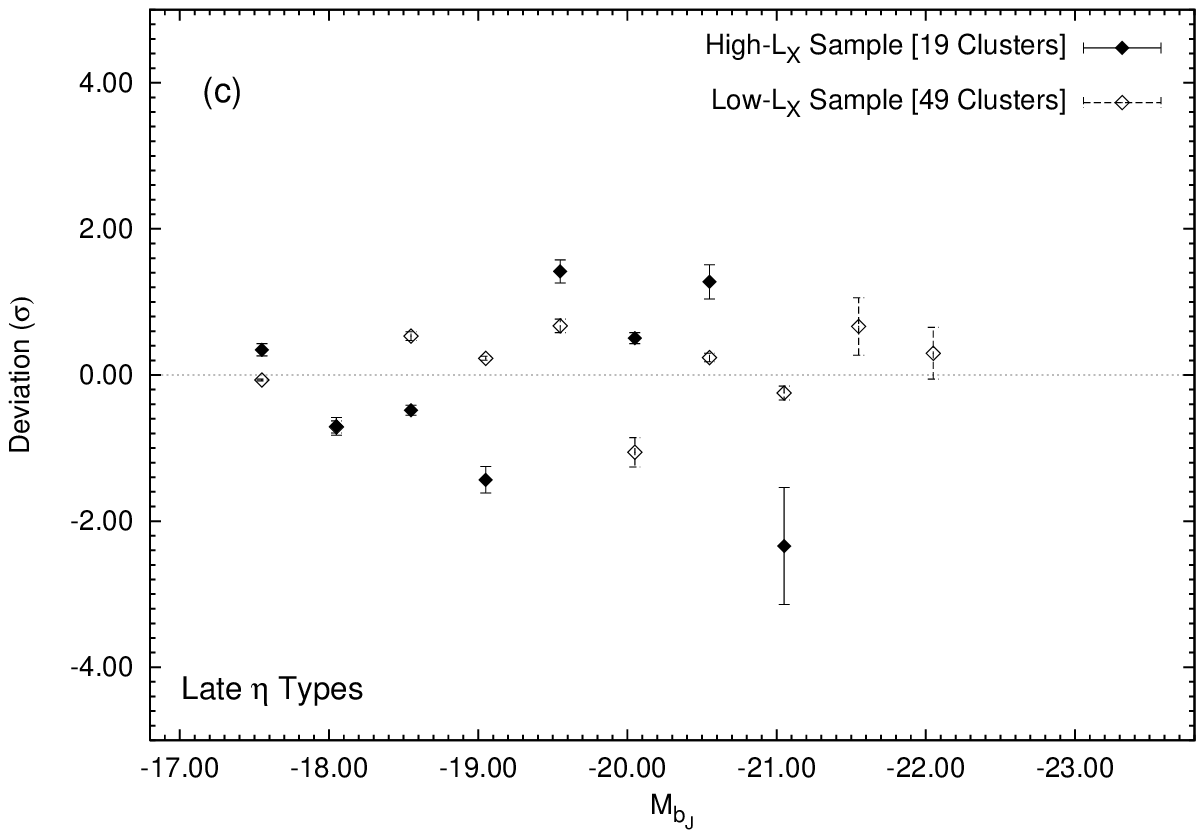}
\includegraphics[width=84mm]{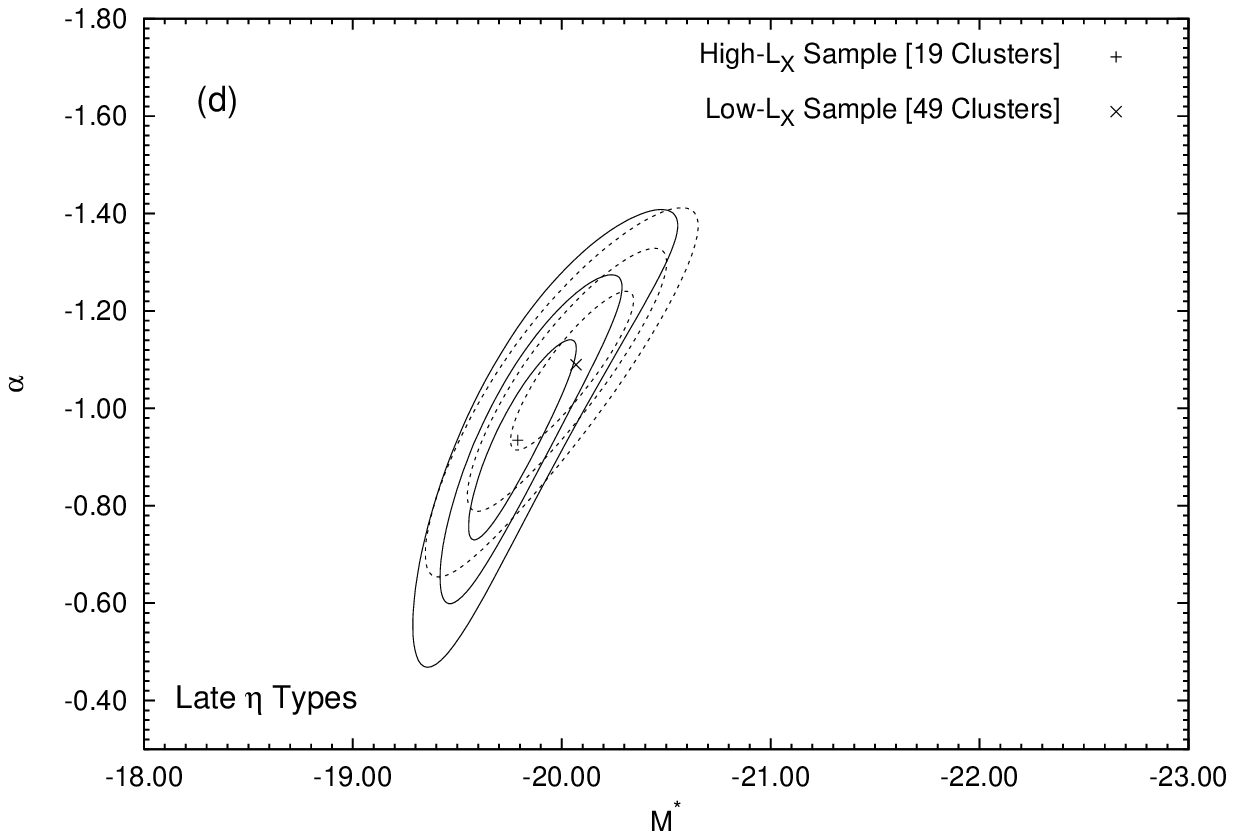}
\caption{Composite luminosity functions for late ($\eta > -1.4$) spectral types for: (a) the high-$L_{\rm X}$ sample and (b) the low-$L_{\rm X}$ sample. The solid line in each plot is the best fitting Schechter function. For comparison, we plot the renormalised fit to the high-$L_{\rm X}$ sample LF as the dashed line in panel (b). Quoted independent errors on each fitted parameter are given at the $1\sigma$ level, and were determined using a bootstrap resampling technique. (c) The deviation of each luminosity function from the best-fitting Schechter function fit to the low-$L_{\rm X}$ sample data, in units of the error on each point. (d) 1, 2, 3$\sigma$ $\chi^2$ error contours for joint luminosity function parameter estimates of the low- and high-$L_{\rm X}$ samples.}
\label{f_LateLFs}
\end{figure*}

The $1\sigma$ errors on the individual Schechter function parameters were estimated using a bootstrap resampling technique, with the other parameter fixed to the value obtained in the $\chi^2$ minimisation. In each bootstrap sample, we forced the number of galaxies drawn from each cluster to be equal to the number of galaxies found in that cluster in the real sample. This ensured that the weighting of each cluster in the production of the composite LF was approximately the same for each bootstrap sample. We believe that the error estimates obtained using this technique are more robust than $\chi^2$ confidence intervals for individual parameters, because by design the bootstrap method takes into account the effect of sample-to-sample variation. We found that the errors estimated using the bootstrap resampling technique were typically $\sim50$ per cent larger than the equivalent $1\sigma$ individual parameter $\chi^2$ confidence intervals.

We now examine the results for each spectral binning in turn.

\subsubsection{All spectral types}
Fig.~\ref{f_AllLFs} shows the resulting composite LFs for galaxies of all spectral types in the low-$L_{\rm X}$ and high-$L_{\rm X}$ samples. Panel (a) of this figure shows that the Schechter function is a poor fit to the high-$L_{\rm X}$ sample LF at the bright-end, where there is a significant excess of galaxies above that predicted by the fitted model. The two brightest bins of this LF contain the Brightest Cluster Galaxies (BCGs) of several clusters, and we found that explicitly removing the BCGs from the sample made no significant difference ($<1\sigma$) to the derived LF parameters. Note that to make a fair comparison with the low-$L_{\rm X}$ LF we have fitted the high-$L_{\rm X}$ LF over the same magnitude range. However, fitting the high-$L_{\rm X}$ LF over the full range in absolute magnitude ($-17.3 > M_{b_J} > -23.3$) does not significantly alter the values of the Schechter function parameters. We obtain $M^*=-20.07\pm0.06$, $\alpha=-0.84\pm0.05$ when the very brightest galaxies are included. In contrast, the Schechter function provides a good fit to the LF of the low-$L_{\rm X}$ clusters along the whole range in absolute magnitude, as can be seen from panel (b). 

In panel (c), we show the difference between the LF value at each point and the best-fitting Schechter function fit for the low-$L_{\rm X}$ sample (after renormalisation to the high-$L_{\rm X}$ LF). The vertical axis in this plot is in units of the error on each LF data point. We can see from this plot that there is a significant deficit of galaxies at $M_{b_J}\sim-21$ in the high-$L_{\rm X}$ sample LF in comparison to the best-fitting Schechter function for the low-$L_{\rm X}$ sample. Panel (d) shows the joint 1, 2, 3$\sigma$ $\chi^2$ error contours on the two-parameter fit. Some overlap of the confidence contours at the $2\sigma$ level is visible in this plot. When we consider the uncertainties on the individual parameter estimates, we find significant differences between the low- and high-$L_{\rm X}$ LFs. $M^*$ is fainter in the high-$L_{\rm X}$ sample than the low-$L_{\rm X}$ sample at $4.7\sigma$ significance ($\Delta M^*=0.51\pm0.11$). The faint-end slope is also shallower in the high-$L_{\rm X}$ sample at $3.3\sigma$ significance ($\Delta \alpha=0.23\pm0.04$).

\begin{table*}
\caption{Comparison of derived LF parameters with the values of \citet{dePropLFs_2003} and \citet{MadgwickLFs_2002}, based on 2dFGRS data. Quoted errors on the individual parameter estimates for this work are $1\sigma$ errors determined using a bootstrap resampling technique.}
\label{t_LFComp}
\begin{tabular}{|c|c|c|c|c|c|}
\hline
Reference			& Spectral type			& $M^{*}$		& $\alpha$	 & Sample\\
\hline
This work -- low-$L_{\rm X}$ 	& All				& -20.57$\pm$0.09	& -1.07$\pm$0.05 & 49 clusters\\
This work -- high-$L_{\rm X}$	& All				& -20.06$\pm$0.06	& -0.84$\pm$0.05 & 19 clusters\\
\citet{dePropLFs_2003}	 	& All				& -20.84$\pm$0.07	& -1.28$\pm$0.03 & 60 clusters\\
\citet{MadgwickLFs_2002}	& All				& -20.56$\pm$0.04	& -1.19$\pm$0.01 & Field\\
\hline
This work -- low-$L_{\rm X}$	& Early				& -20.70$\pm$0.12	& -0.97$\pm$0.06 & 49 clusters\\
This work -- high-$L_{\rm X}$	& Early				& -20.12$\pm$0.08 	& -0.79$\pm$0.06 & 19 clusters\\
\citet{dePropLFs_2003}	 	& Type 1			& -20.81$\pm$0.09 	& -1.05$\pm$0.04 & 60 clusters\\
\citet{MadgwickLFs_2002}	& Type 1			& -20.35$\pm$0.05 	& -0.54$\pm$0.02 & Field\\
\hline
This work -- low-$L_{\rm X}$	& Late				& -20.07$\pm$0.14 	& -1.09$\pm$0.08 & 49 clusters\\
This work -- high-$L_{\rm X}$	& Late				& -19.79$\pm$0.11 	& -0.93$\pm$0.10 & 19 clusters\\
\citet{dePropLFs_2003}	 	& Type 2			& -20.25$\pm$0.13 	& -1.23$\pm$0.07 & 60 clusters\\
\citet{dePropLFs_2003}	 	& Types 3+4			& -19.91$\pm$0.19	& -1.30$\pm$0.10 & 60 clusters\\
\citet{MadgwickLFs_2002}	& Type 2			& -20.30$\pm$0.03 	& -0.99$\pm$0.01 & Field\\
\citet{MadgwickLFs_2002}	& Type 3			& -19.94$\pm$0.04 	& -1.24$\pm$0.02 & Field\\
\citet{MadgwickLFs_2002}	& Type 4			& -19.92$\pm$0.05 	& -1.50$\pm$0.03 & Field\\
\citet{MadgwickLFs_2002}	& Late (weighted average)	& -20.11$\pm$0.04 	& -1.16$\pm$0.02 & Field\\
\hline
\end{tabular}
\end{table*}

\subsubsection{Early spectral types}
Because both cluster samples are dominated by passively evolving galaxies, the results for the early-type LFs shown in Fig.~\ref{f_EarlyLFs} follow a similar pattern to that for the LFs of all spectral types. We see again that the Schechter function is a poor fit to the high-$L_{\rm X}$ sample LF at the bright-end, and that the low-$L_{\rm X}$ sample is well fitted by the Schechter function over its entire range. Again, a fit to the high-$L_{\rm X}$ LF over the full range in absolute magnitude ($-17.3 > M_{b_J} > -23.3$) does not give significantly different results from those obtained for the range ($-17.3 > M_{b_J} > -22.3$) -- we obtain $M^*=-20.14\pm0.08$, $\alpha=-0.80\pm0.06$. $M^*$ and $\alpha$ are both found to be lower in the high-$L_{\rm X}$ clusters compared to the low-$L_{\rm X}$ sample.

Inspection of panel (d) shows a small amount of overlap between the $2\sigma$ confidence contours for the joint parameter estimates. If we consider the uncertainties on the individual parameters, the difference in the LF characteristic magnitudes is significant at the $4\sigma$ level ($\Delta M^*=0.58\pm0.14$). In contrast, the difference between the $\alpha$ values is only significant at the $2.3\sigma$ level ($\Delta \alpha=0.18\pm0.08$). Panel (c) shows the main reason for the lower value of $M^*$ in the Schechter function fit to the high-$L_{\rm X}$ sample LF -- there is a significant ($\sim 4\sigma$) deficit of early spectral type galaxies with $M_{b_J}\sim-21$ in comparison to the best Schechter function fit to the low-$L_{\rm X}$ sample LF.

\subsubsection{Late spectral types}
In panels (a) and (c) of Fig.~\ref{f_LateLFs}, we see that the high-$L_{\rm X}$ LF contains no galaxies with $M_{b_J}<-21.3$. Despite this, panel (d) shows that there are no significant differences between the LF parameter values obtained for each sample when the confidence contours for the joint parameter fit are considered. The uncertainties in the individual LF parameters indicate likewise: $\Delta M^*=0.28\pm0.18$, $\Delta \alpha=0.16\pm0.13$. Therefore the differences in the LF parameters found for galaxies of all spectral types are caused by the different LFs of early-type, passively evolving galaxies.

\section{Discussion}
\label{s_Discuss}
\subsection{Early-type galaxy fractions}
In Section~\ref{s_properties} we established that in both the low- and high-$L_{\rm X}$ cluster samples the early spectral type galaxy fraction falls steadily with increasing distance from the cluster centre (Fig.~\ref{f_Radial}). Although the 2dFGRS spectral type parameter $\eta$ is correlated with morphology with a large amount of scatter, this behaviour is reminiscent of the long-established galaxy morphology--local density relation in clusters \citep[e.g.][]{Dressler_1980}. In Fig.~\ref{f_Radial}, we also see that the fraction of passively evolving, early spectral type galaxies is higher in the high-$L_{\rm X}$ sample than in the low-$L_{\rm X}$ sample. \citet{BaloghHSTa_2002} reported a result similar in character to that found in this study from a morphological study of low- and high-$L_{\rm X}$ clusters conducted with the Hubble Space Telescope. They found that within a fixed physical region, the mean fraction of disc-dominated galaxies is strongly dependent upon $L_{\rm X}$. Although the epoch of this study was different ($z\sim0.25$ compared to $z<0.1$ for our study), as are the observations and quantities measured, the fact that spectral type is correlated with morphology, albeit with large scatter \citep{MadgwickEta_2003}, means that a comparison of our result with one of the conclusions of \citet{BaloghHSTa_2002} is not unreasonable. \citet{BaloghHSTa_2002} interpreted their results as indicating that galaxy mergers have played a bigger role in the evolution of more massive clusters.

\begin{table*}
\caption{Luminosity function parameters derived for studies using SDSS $g$-band photometric data, transformed to $H_0=70$ km s$^{-1}$ Mpc$^{-1}$ and the 2dFGRS $b_J$ band.}
\label{t_SDSSLFComp}
\begin{tabular}{|c|c|c|c|c|c|}
\hline
Reference			& Spectral type			& $M^{*}$		& $\alpha$	 & Sample\\
\hline
\citet{PopessoII_2005} & All	& -21.11$\pm$0.20 & -1.33$\pm$0.04 & 69 clusters (members selected within 1.5 Mpc radius)\\
\citet{PopessoIVpre_2005} & All & -20.70$\pm$0.21 & -1.07$\pm$0.12 & 69 clusters (members selected within $R_{200}$)\\
\citet{Blanton_2003} & All & -19.91$\pm$0.02 & -0.89$\pm$0.03 & Field\\
\hline
\citet{PopessoIVpre_2005} & Early & -20.31$\pm$0.16 & -0.69$\pm$0.10 & 15 clusters (members selected within $R_{200}$)\\
\hline
\citet{PopessoIVpre_2005} & Late & -21.65$\pm$0.40 & -1.80$\pm$0.04 & 15 clusters (members selected within $R_{200}$)\\
\hline
\end{tabular}
\end{table*}

Since both $L_{\rm X}$ and $\sigma_r$ should trace the cluster mass, and $\eta$ is correlated with the equivalent width of the H$\alpha$ line, we can also make a comparison with the results of \citet{Balogh2dFSDSS_2004}, who examined the correlation between the fraction of galaxies with H$\alpha$ equivalent width $>4$\AA \ and local density in the 2dFGRS and SDSS. \citet{Balogh2dFSDSS_2004} divided a sample of $\sim 25,000$ group and cluster galaxies into two subsamples of low-$\sigma_r$ ($200 < \sigma_r < 400$ km s$^{-1}$) and high-$\sigma_r$ ($500 < \sigma_r <1000$ km s$^{-1}$). They found marginal evidence that the fraction of star-forming galaxies was lower in the high-$\sigma_r$ clusters relative to the low-$\sigma_r$ groups at fixed local density. This is similar to the result we see for the low- and high-$L_{\rm X}$ cluster samples studied in this paper, although we make no attempt to distinguish between the effects of variation in local galaxy density and the large scale cluster environment (as traced by $L_{\rm X}$). \citet{Lewis2dF_2002} also studied star-formation rates in 2dFGRS cluster galaxies as traced by the equivalent width of H$\alpha$. \citet {Lewis2dF_2002} considered 17 clusters drawn from DP02 and divided their sample into clusters with $\sigma_r >800$ km s$^{-1}$ and $\sigma_r < 800$ km s$^{-1}$, with 10 and 7 clusters in each respective subsample. They found no dependence of galaxy star-formation rates upon the velocity dispersion of the host cluster, in contrast to the results that we obtain for our low- and high-$L_{\rm X}$ cluster samples. This is perhaps due to the comparatively small size of the cluster sample used by \citet{Lewis2dF_2002}.

\subsection{Luminosity functions}
We present a comparison of our derived LF parameters with the values obtained from 2dFGRS data by DP03 for clusters and \citet{MadgwickLFs_2002} for the field in Table~\ref{t_LFComp}, divided by spectral type and transformed to $H_0=70$ km s$^{-1}$ Mpc$^{-1}$. Here, spectral type 1 corresponds to our definition of early-type galaxies and spectral types 2-4 collectively form our sample of late-type galaxies. In this table we also quote average values of $M^*$ and $\alpha$ for the late-type LFs of \citet{MadgwickLFs_2002}, weighted by the respective normalisation constants, to ease comparison with our late-type LF parameter values.

\begin{figure*}
\includegraphics[width=70mm]{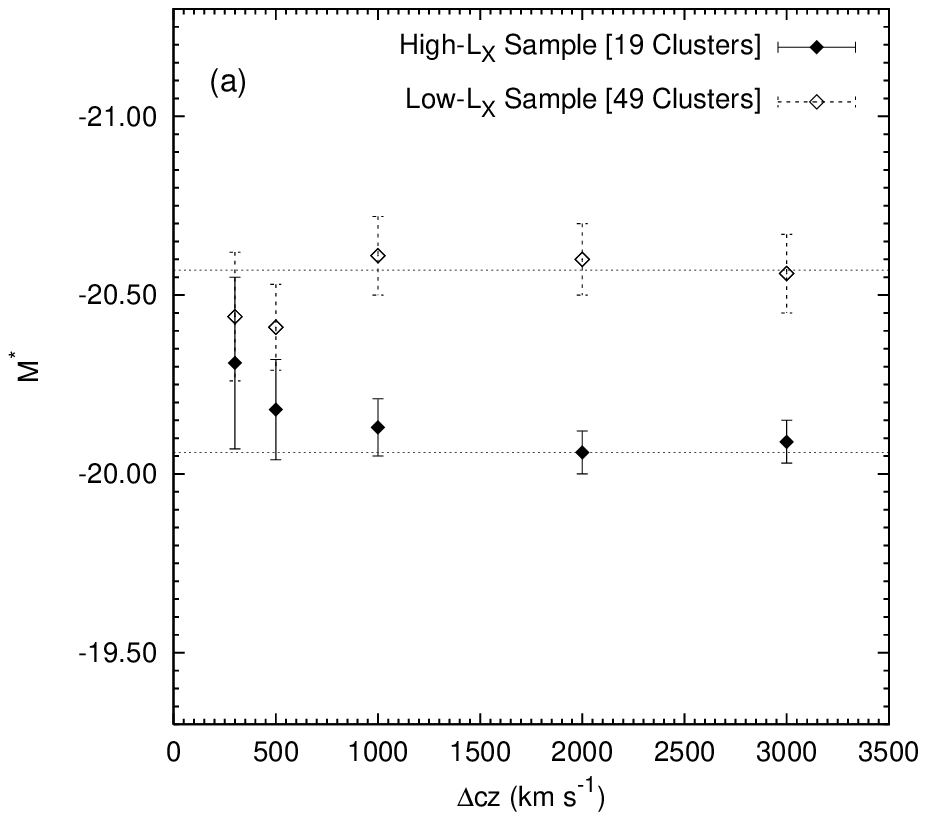}\includegraphics[width=70mm]{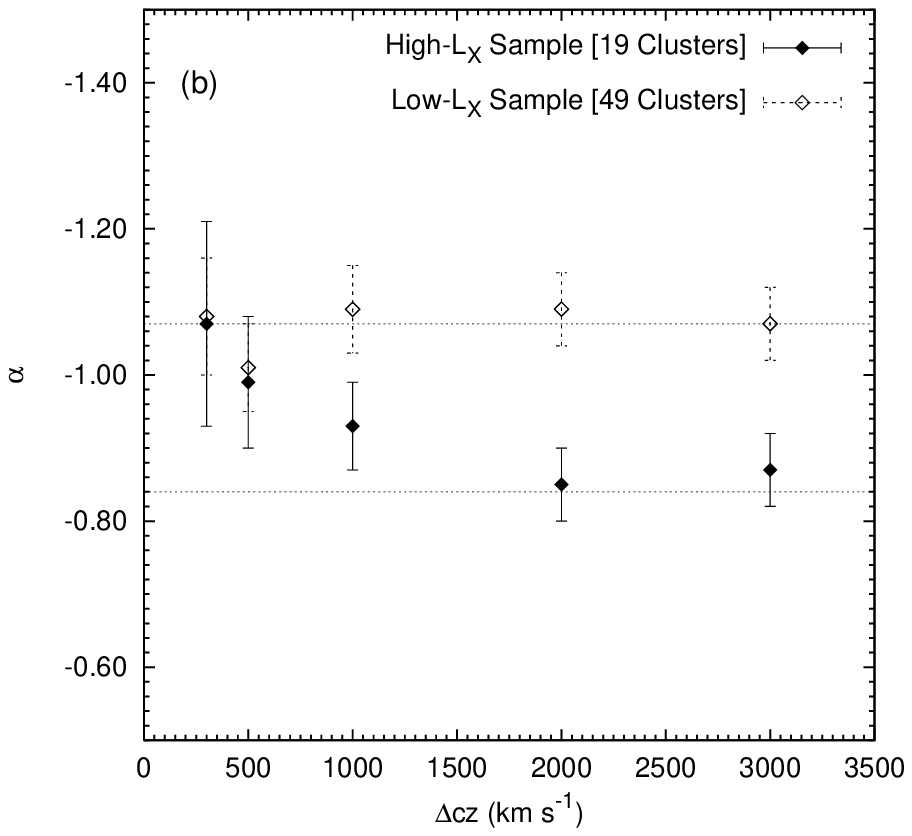}
\caption{The effect on (a) $M^*$ and (b) $\alpha$ of limiting selection in redshift space to different slices $\Delta cz$ around each cluster redshift, for the luminosity functions of galaxies of any spectral type. The dotted lines show the best-fit value for the appropriate LF parameter and sample obtained using the $3\times\sigma_r$ selection in $cz$.}
\label{f_SBISlices}
\end{figure*}

The most surprising result of this work is the disagreement of the high-$L_{\rm X}$ sample early-type LF parameter values with both the low-$L_{\rm X}$ sample and the results of DP03. $M^*$ and $\alpha$ are both lower in the high-$L_{\rm X}$, early-type LF in comparison to the corresponding DP03 early-type LF at $>3\sigma$ significance. The characteristic magnitude of the high-$L_{\rm X}$ sample LF is also significantly lower than found in the field by \citet{MadgwickLFs_2002}, although the faint-end slope is steeper. In contrast, both Schechter function parameter values for the early-type LF of the low-$L_{\rm X}$ sample are within $\sim1\sigma$ of agreement with the results of DP03, with $M^*$ brighter than in the field, and steeper $\alpha$. 

The late-type LF parameter values for the high-$L_{\rm X}$ sample agree within $<3\sigma$ of the type 2 and types 3+4 LFs of DP03, and the weighted average of the field results of \citet{MadgwickLFs_2002}. There is therefore marginal disagreement -- although note that there is no significant disagreement between the low- and high-$L_{\rm X}$ late-type LFs. The low-$L_{\rm X}$ sample late-type LF parameters are well within $<2\sigma$ of agreement with the DP03 results and the \citet{MadgwickLFs_2002} field results. These results are therefore consistent with a picture in which the late-type LF varies little with environment.

We are also able to compare our results to similar studies undertaken using SDSS $g$-band data, following a suitable transformation to the $b_J$ band as used by the 2dFGRS, i.e.,

\begin{equation}
\label{e_g2bj}
b_J=g+0.15+0.13(g-r),
\end{equation}
as used by \citet{dePropBO_2004}. We present results from \citet{PopessoII_2005, PopessoIVpre_2005} for RASS-SDSS galaxy clusters, and \citet{Blanton_2003} for the field in Table~\ref{t_SDSSLFComp}. Note that we assume a $(g-r)$ colour of 0.8 when converting the $M^*$ values using equation~(\ref{e_g2bj}), a value we estimated from inspection of the colour distribution of early spectral type 2dFGRS galaxies, after transforming the 2dFGRS $b_J$ and $R_F$ bands to Sloan $g$ and $r$ using the prescriptions given in \citet{dePropBO_2004}.

\citet{PopessoII_2005} computed LFs for a sample of 69 X-ray selected clusters drawn from the RASS-SDSS survey. We quote the LF parameter values for the composite LF produced from galaxies within 1.5 Mpc clustercentric distance in Table~\ref{t_SDSSLFComp}. We can see from comparison of the results presented in Section~\ref{s_results} that there is a significant disagreement between \citet{PopessoII_2005} and the results that we obtain for both LF parameters for the high-$L_{\rm X}$ sample LF of all spectral types -- in fact, we obtain a value of $M^*$ that is $\sim 1$ magnitude fainter. In contrast, there is mild disagreement (at the $\sim2.5\sigma$ level) between the characteristic magnitude of this LF and the value that we obtain for the low-$L_{\rm X}$ sample LF for galaxies of all spectral types. However, our low-$L_{\rm X}$ sample LF has a significantly flatter faint-end slope than this LF. 

\citet{PopessoIVpre_2005} constructed LFs for the same 69 RASS-SDSS clusters using galaxies selected within $R_{200}$, rather than using a fixed metric aperture, and fitted a composite of two Schechter functions to each LF. Using this method, they obtained results consistent with those previously obtained in \citet{PopessoII_2005}. It is appropriate to compare the LF values that we obtained in Section~\ref{s_results} with the bright component of the \citet{PopessoIVpre_2005} double-Schechter function fits. We find in this case a much better agreement of our low-$L_{\rm X}$ sample LF for galaxies of all spectral types, with the values of $M^*$ and $\alpha$ in agreement within $\sim 1 \sigma$ of each other. The agreement with the LF parameters we derived for the equivalent high-$L_{\rm X}$ sample LF is also much closer -- we find $\Delta M^*=0.64\pm0.22$ and $\Delta \alpha=0.23\pm0.13$ in this case. 

\citet{PopessoIVpre_2005} also constructed LFs for galaxies divided into early and late types on the basis of their $u-r$ colour.
Because of uncertainties in the $K$-correction, they used a reduced sample of sixteen RASS-SDSS clusters located within $z<0.1$. As Table~\ref{t_SDSSLFComp} shows, \citet{PopessoIVpre_2005} found a value of $M^*$ for the early-type galaxy LF intermediate between the values we obtained for our low- and high-$L_{\rm X}$ cluster samples. In fact, both $M^*$ and $\alpha$ for this LF are in good agreement with the values we obtain for our corresponding high-$L_{\rm X}$ sample LF. The agreement with the low-$L_{\rm X}$ sample LF is worse, but the $M^*$ values are still in agreement within $<2\sigma$, and the $\alpha$ values differ at the $\sim2.3\sigma$ level. However, for the late-type LF \citet{PopessoIVpre_2005} find a much brighter characteristic magnitude and steeper faint-end slope than we find for both our low- and  high-$L_{\rm X}$ cluster samples. We note that unlike the late-type LFs produced from 2dFGRS data by \citet{MadgwickLFs_2002} for the field, and both DP03 and the present work for clusters, \citet{PopessoIVpre_2005} obtain a brighter characteristic magnitude for the late-type LF in comparison to the early-type LF. The reasons for this discrepancy are not clear.

It is prudent to look for selection effects that could explain the discrepancy between the early-type luminosity functions in our low- and high-$L_{\rm X}$ cluster samples. In the case of the low-$L_{\rm X}$ sample, which are almost all exclusively optically selected, we may worry that they are not in fact bound systems, in which case the effect of interloper galaxies on the derived LF parameters could be significant. To investigate the effect of changing the selection of cluster members in redshift space on the derived LF parameters, we constructed LFs using galaxies selected within $\pm300$, $\pm500$, $\pm1000$, $\pm2000$, and $\pm3000$ km s$^{-1}$ slices ($\Delta cz$) around each cluster redshift, within the previously calculated selection radius of $R_{200}$ (i.e. instead of using $3 \times$ the velocity dispersion to select cluster members). Fig.~\ref{f_SBISlices} shows the resulting Schechter function parameters, with 1$\sigma$ statistical errors derived from bootstrap resampling, obtained using the same method as the results presented in Section~\ref{s_LFs}. We can see from this figure that the values of $M^*$ and $\alpha$ are not significantly affected by this selection effect, remaining consistent with the results presented in Section~\ref{s_results}.

\begin{figure*}
\includegraphics[width=84mm]{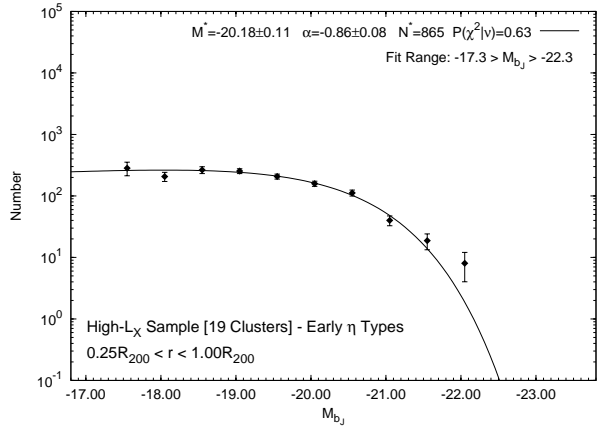}
\caption{The luminosity function for early ($\eta\leq-1.4$) spectral types for the high-$L_{\rm X}$ cluster sample, after excising galaxies within $0.25R_{200}$ of cluster centres.}
\label{f_coreless}
\end{figure*}

We also considered the possibility that a selection effect could arise due to galaxy crowding in the centres of the high-$L_{\rm X}$ clusters, because these objects are on average more distant than the low-$L_{\rm X}$ clusters (panel (b) of Fig.~\ref{f_SBIDist}). There are two possible sources of unaccounted-for incompleteness in this case: incompleteness in the parent APM galaxy catalogue, due to object blending; and/or incompleteness in the 2dFGRS spectroscopic catalogue, due to fibre crowding. To determine the importance of this effect, we produced luminosity functions for the high-$L_{\rm X}$ sample after excising galaxies within the central $0.25R_{200}$ of each cluster. We present the LF for early spectral types in Fig.~\ref{f_coreless}. The derived LF parameters ($M^*=-20.18\pm0.11$, $\alpha=-0.86\pm0.08$) are clearly consistent with those obtained in Section~\ref{s_results}, and so we conclude that incompleteness effects of this type have no significant impact upon our results. 

The physical aperture sampled by a 2dF fibre naturally becomes smaller as redshift increases, and so the spectral type determined for a particular galaxy may have a mild redshift dependence. However, due to seeing effects and fibre positioning errors, the effect is likely to be small \citep[see][]{MadgwickLFs_2002}. Nevertheless, we attempted to gauge the impact of the aperture bias by constructing LFs for both the low- and high-$L_{\rm X}$ samples using only clusters located within a common redshift range of $0.03 < z < 0.07$. Of the high-$L_{\rm X}$ cluster sample, only 8 clusters of the 19 are found in this redshift range, and we found the best fitting early-type LF parameters were $M^*=-19.90\pm0.13$, $\alpha=-0.74\pm0.08$. The low-$L_{\rm X}$ sample in this redshift range is made up of 39 clusters, for which the best fitting Schechter function was specified by $M^*=-20.76\pm0.13$, $\alpha=-0.96\pm0.06$. We therefore conclude from a comparison of these results with those for the full sample that the effect of fibre aperture bias is negligible. 

\begin{figure*}
\includegraphics[width=84mm]{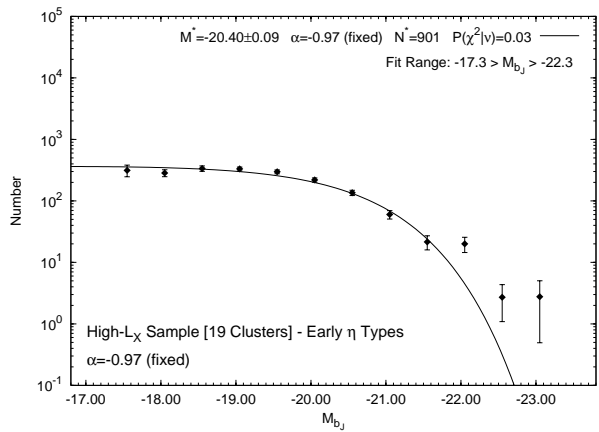}
\caption{Single parameter fit to the luminosity function for early ($\eta\leq-1.4$) spectral types for the high-$L_{\rm X}$ cluster sample. The faint-end slope has been fixed to $\alpha=-0.97$, the value found from a fit to the corresponding low-$L_{\rm X}$ sample LF.}
\label{f_fixedAlpha}
\end{figure*}

The faint-end of the high-$L_{\rm X}$ sample is defined by relatively few clusters. In fact, at magnitudes fainter than $M_{b_J}>-18.8$, only 11 clusters contribute to the LF. Although the early-type high-$L_{\rm X}$ sample LF is made up from 17 per cent more galaxies than the equivalent low-$L_{\rm X}$ LF, we may simply be unfortunate in that the faint-end behaviour of these 11 clusters is not typical of the rest of the sample. To address this concern, we performed a one-parameter fit to the early-type, high-$L_{\rm X}$ sample LF, with the value of the faint-end slope fixed to that found for the low-$L_{\rm X}$ sample, i.e. $\alpha=-0.97$. The resulting LF is shown in Fig.~\ref{f_fixedAlpha}. Although the value of $M^*$ for the best fit is now brighter than found from the two-parameter fit, we find that $M^*$ remains lower than the value obtained from the low-$L_{\rm X}$ sample LF. This result is marginally significant, at the $2\sigma$ level ($\Delta M^*=0.30\pm0.15$). Nevertheless, this exercise confirms that the low-value of $M^*$ found for the high-$L_{\rm X}$ cluster sample arises primarily from the deficit of early-type galaxies with $M_{b_J}\sim-21$ (panel (c) of Fig.~\ref{f_EarlyLFs}), rather than from the behaviour of the faint-end of the LF.
 
\subsection{Interpretation}
Since simple selection effects are unable to explain our results, we now consider how the difference in the early-type LF parameters could be driven by environmental differences between the low- and high-$L_{\rm X}$ clusters. Other studies of the luminosity function using 2dFGRS data have found evidence to suggest that $M^*$ brightens and $\alpha$ steepens with increasing galaxy density \citep{Croton_2005} or mass \citep{Eke_2004}. The results presented here seem to suggest that the opposite is true when considering the effect of increasing X-ray luminosity on the LF. It could therefore be suggested that the high-$L_{\rm X}$ cluster sample is actually composed of low-mass, merging systems -- $L_{\rm X}$ is known to be a good tracer of mass, but the results of numerical simulations suggest that $L_{\rm X}$ can be boosted during mergers. 

For example, \citet*[][see also \citealt{RickerSarazin_2001}]{RandallMergers_2002} demonstrated that an equal mass merger with impact parameter $b=0$ (i.e. a head-on collision) can boost the bolometric X-ray luminosity of the resultant object by a factor of 5-6. The boost in $L_{\rm X}$ in this case lasts for approximately half a sound crossing time, $\sim 1$ Gyr for the high-$L_{\rm X}$ cluster sample. Without knowledge of the merger histories of the individual high-$L_{\rm X}$ clusters, the impact of this effect is difficult to gauge. However, in the limiting case of all the high-$L_{\rm X}$ clusters being boosted in $L_{\rm X}$ by a factor of six, the median unboosted (0.1--2.4 keV) X-ray luminosity of the sample is $0.16 \times 10^{44}$ erg s$^{-1}$. By comparison, for the majority optically-selected low-$L_{\rm X}$ cluster sample the median maximum $L_{\rm X}$ set by the REFLEX flux limit is $0.22 \times 10^{44}$ erg s$^{-1}$. Therefore even in this extreme case both samples would be of similar mass -- rather than the high-$L_{\rm X}$ clusters being significantly less massive. In addition, as shown in Section~\ref{s_sampleDefinitions}, the median velocity dispersion of the high-$L_{\rm X}$ sample is higher than that for the low-$L_{\rm X}$ sample. These facts imply that the high-$L_{\rm X}$ clusters are more likely to be higher mass systems than the low-$L_{\rm X}$ clusters.

We suggest that a consistent explanation of the results of Sections~\ref{s_properties} and~\ref{s_results} can be achieved by a scenario in which the high-$L_{\rm X}$ clusters are in general older, more dynamically evolved systems than the low-$L_{\rm X}$ clusters. In this case, the higher fraction of early-type, passively evolving galaxies present in the high-$L_{\rm X}$ sample comes about naturally from there being more crossing times available for infalling late-type galaxies to be converted into early-type galaxies within the cluster environment. This holds irrespective of the mechanism responsible for driving the transformation, whether it is by interactions with the intracluster medium (as in ram-pressure stripping), or by interactions between cluster member galaxies (as in the harassment scenario). 

The excess galaxies that we see at the very bright end of the high-$L_{\rm X}$ cluster LF could be explained by these systems having a more extensive merger history than the low-$L_{\rm X}$ clusters, as suggested by the results of \citet{BaloghHSTa_2002}. In particular, a `galactic cannibalism' scenario \citep[e.g.][]{HausmanOstriker_1978} -- where the brightest cluster galaxies grow by the accretion of the successively fainter members -- could help to bring about the deficit of galaxies seen at $M_{b_J}\sim-21$ in the high-$L_{\rm X}$ sample, early-type LF, which seems to be the cause of the low-value of $M^*$ found in this case. 

\citet*{LinMohrStanford_2004} studied the $K$-band LFs of 93 galaxy clusters divided into two samples of low- and high-mass (estimated from the X-ray temperature). Although they found the high-mass LF had a brighter characteristic magnitude than the low-mass LF -- in contrast to the results of this work for low- and high-$L_{\rm X}$ clusters -- they found a deficit of galaxies below $M^*$ in the high-mass sample in comparison to the low-mass LF. \citet{LinMohrStanford_2004} \citep[see also][]{LinMohr_2004} interpreted their result as an indication that a major contribution to the masses of the very brightest galaxies in clusters comes from mergers with other bright ($\sim M^*$) cluster galaxies. Perhaps because our LFs are constructed purely from spectroscopically-selected cluster members -- rather than employing a more uncertain statistical subtraction of non-cluster members -- the effect that we see on the derived Schechter function parameters is more dramatic. 

\section{Conclusions}
\label{s_Conclude}
We have correlated the REFLEX catalogue of X-ray selected galaxy clusters with the 2dF Galaxy Redshift Survey and performed a study of the dependence of cluster galaxy populations upon X-ray luminosity. Using this large, homogeneous combined dataset, supplemented with additional rich clusters from the catalogue compiled by \citet{dePropClusters_2002}, we found:

(1) An $L_{\rm X}-\sigma_r$ relation consistent with that derived by \citet{Ortiz-Gil_2004} for the REFLEX survey team using a combination of literature and REFLEX optical follow-up data.

(2) The fraction of early-type, passively evolving galaxies is 12 per cent higher in high-$L_{\rm X}$ clusters compared to low-$L_{\rm X}$ clusters out to $R_{200}$, as determined from a volume-limited sample of 2dFGRS galaxies with $M_{b_J}<-19$, $z<0.11$. In both cases the early-type galaxy fraction falls off smoothly with increasing distance from the cluster centre.

(3) Both the characteristic magnitude $M^{*}$ and faint-end slope $\alpha$ of the galaxy luminosity function are lower in high-$L_{\rm X}$ clusters compared to low-$L_{\rm X}$ clusters. This is caused primarily by the low value of $M^*$ found for the LF of early spectral type galaxies in the high-$L_{\rm X}$ sample LF, which is driven by an underabundance of galaxies with $M_{b_J}\sim-21$ relative to the corresponding low-$L_{\rm X}$ sample LF. 

We believe these results are consistent with a scenario in which the high-$L_{\rm X}$ clusters are more dynamically evolved systems than the low-$L_{\rm X}$ clusters. The higher fraction of passively evolving galaxies found in the high-$L_{\rm X}$ sample could arise through the accretion and conversion of a greater number of star-forming field galaxies. The lower value of $M^*$ in the LF of early spectral types galaxies in high-$L_{\rm X}$ clusters could be explained by `cannibalism' of bright ($M_{b_J}\sim-21$) members to produce the very brightest cluster galaxies.

\section*{Acknowledgments}
We thank everyone involved in the success of both the 2dF Galaxy Redshift Survey and the REFLEX cluster survey, without which of course this work would not have been possible. Thanks to Phil James for some useful suggestions that improved the clarity of this paper. Thanks also to the referee, Hans B\"ohringher, who provided useful feedback which improved this manuscript. M. Hilton acknowledges the support of a studentship from Liverpool John Moores University.

\setlength{\bibhang}{2.0em}

\label{lastpage}

\end{document}